\documentstyle[12pt,epsfig]{article}

\catcode`\@=11
\def\@citex[#1]#2{%
\if@filesw \immediate \write \@auxout {\string \citation {#2}}\fi
\@tempcntb\m@ne \let\@h@ld\relax \def\@citea{}%
\@cite{%
  \@for \@citeb:=#2\do {%
    \@ifundefined {b@\@citeb}%
      {\@h@ld\@citea\@tempcntb\m@ne{\bf ?}%
      \@warning {Citation `\@citeb ' on page \thepage \space undefined}}%
      {\@tempcnta\@tempcntb \advance\@tempcnta\@ne%
      \@tempcntb\number\csname b@\@citeb \endcsname \relax%
      \ifnum\@tempcnta=\@tempcntb 
        \ifx\@h@ld\relax%
          \edef \@h@ld{\@citea\csname b@\@citeb\endcsname}%
        \else%
          \edef\@h@ld{\ifmmode{-}\else--\fi\csname b@\@citeb\endcsname}%
        \fi%
      \else
        \@h@ld\@citea\csname b@\@citeb \endcsname%
        \let\@h@ld\relax%
      \fi}%
    \def\@citea{,\penalty\@highpenalty\,}%
  }\@h@ld
}{#1}}

\def\@citeb#1#2{{[#1]\if@tempswa , #2\fi}}
\def\@citeu#1#2{{$^{#1}$\if@tempswa , #2\fi }}
\def\@citep#1#2{{#1\if@tempswa , #2\fi}}

\def\bcites{         
        \catcode`\@=11
        \let\@cite=\@citeb
        \catcode`\@=12
}

\def\upcites{         
        \catcode`\@=11
        \let\@cite=\@citeu
        \catcode`\@=12
}

\def\plaincites{      
        \catcode`\@=11
        \let\@cite=\@citep
        \catcode`\@=12
}

\newcount\hour
\newcount\minute
\newtoks\amorpm
\hour=\time\divide\hour by 60
\minute=\time{\multiply\hour by 60 \global\advance\minute by-\hour}
\edef\standardtime{{\ifnum\hour<12 \global\amorpm={am}%
        \else\global\amorpm={pm}\advance\hour by-12 \fi
        \ifnum\hour=0 \hour=12 \fi
        \number\hour:\ifnum\minute<10 0\fi\number\minute\the\amorpm}}
\edef\militarytime{\number\hour:\ifnum\minute<10 0\fi\number\minute}

\def\draftlabel#1{{\@bsphack\if@filesw {\let\thepage\relax
   \xdef\@gtempa{\write\@auxout{\string
      \newlabel{#1}{{\@currentlabel}{\thepage}}}}}\@gtempa
   \if@nobreak \ifvmode\nobreak\fi\fi\fi\@esphack}
        \gdef\@eqnlabel{#1}}
\def\@eqnlabel{}
\def\@vacuum{}
\def\marginnote#1{}
\def\draftmarginnote#1{\marginpar{\raggedright\scriptsize\tt#1}}
\def\draft{
        \pagestyle{plain}
        \overfullrule=2pt
        \oddsidemargin -.5truein
        \def\@oddhead{\sl \phantom{\today\quad\militarytime} \hfil
        \smash{\Large\sl DRAFT} \hfil \today\quad\militarytime}
        \let\@evenhead\@oddhead
        \let\label=\draftlabel
        \let\marginnote=\draftmarginnote
        \def\ps@empty{\let\@mkboth\@gobbletwo
        \def\@oddfoot{\hfil \smash{\Large\sl DRAFT} \hfil}
        \let\@evenfoot\@oddhead}
        \def\@eqnnum{(\theequation)\rlap{\kern\marginparsep\tt\@eqnlabel}%
        \global\let\@eqnlabel\@vacuum}  }

\def\blackfonts{
        \font\blackboard=msbm10 scaled\magstep1
        \font\blackboards=msbm8
        \font\blackboardss=msbm6
}

\def\nblack{            
        \def\ZZ{{Z \n{10} Z}}
        \def\NN{{N \n{14} N}}
        \def\CC{{C \n{11} C}}
        \def\RR{{R \n{11} R}}
        \def\QQ{{Q \n{12} Q}}
        \def\PP{{P \n{11} P}}
}

\def\prep{         
        \catcode`\@=11
        \input art10.sty
        \catcode`\@=12
        
        \let\small\null
        \def\blackfonts{
                \font\blackboard=msbm10
                \font\blackboards=msbm7
                \font\blackboardss=msbm5
        }
        \let\sl\it
        \twocolumn
        \sloppy
        \voffset=-2.54truecm
        \hoffset=-2.54truecm
        \flushbottom
        \parindent 1em
        \leftmargini 2em
        \leftmarginv .5em
        \leftmarginvi .5em
        \marginparwidth 48pt
        \marginparsep 10pt
        \setlength{\columnsep}{2truecm}
        \setlength{\textwidth}{25.4truecm}
        \setlength{\textheight}{17truecm}
        \baselineskip=16pt
        \oddsidemargin .18truein
        \evensidemargin .17truein
}

\def\eqalign#1{\null\,\vcenter{\openup\jot\m@th
  \ialign{\strut\hfil$\displaystyle{##}$&$\displaystyle{{}##}$\hfil
      \crcr#1\crcr}}\,}
\def\eqalignno#1{\displ@y \tabskip\centering
  \halign to\displaywidth{\hfil$\@lign\displaystyle{##}$\tabskip\z@skip
    &$\@lign\displaystyle{{}##}$\hfil\tabskip\centering
    &\llap{$\@lign##$}\tabskip\z@skip\crcr
    #1\crcr}}

\def\section{\@startsection {section}{1}{\z@}{3.ex plus 1ex minus
 .2ex}{2.ex plus .2ex}{\large\bf}}
\def\subsection{\@startsection{subsection}{2}{\z@}{2.75ex plus 1ex minus
 .2ex}{1.5ex plus .2ex}{\bf}}        

\def\appendix{{\newpage\section*{Appendix}}\let\appendix\section%
        {\setcounter{section}{0}
        \gdef\thesection{\Alph{section}}}\section}

\def\abstract{\if@twocolumn
\section*{Abstract}
\else 
\begin{center}
{\bf Abstract\vspace{-.5em}\vspace{0pt}}
\end{center}
\quotation
\fi}

\catcode`\@=12

\newcommand{\beq}{\begin{equation}}
\newcommand{\eeq}{\end{equation}}
\newcommand{\beqa}{\begin{eqnarray}}
\newcommand{\eeqa}{\end{eqnarray}}

\newcommand{\Z}{{\bf Z}}

\newcommand{\R}{{\bf R}}
\newcommand{\C}{{\bf C}}
\newcommand{\e}{{\rm e}}

\newcommand{\tilQ}{\widetilde{Q}}

\newcommand{\NSp}{NS${}^{\prime}$\,}
\newcommand{\mua}{\mu_{\rm a}}

\newcommand{\dd}{{\rm d}}
\newcommand{\elst}{{\ell_{\it st}}}
\newcommand{\elel}{{\ell_{11}}}
\newcommand{\gst}{{g_{\it st}}}
\newcommand{\MT}{{$M$ theory}~}
\newcommand{\tilNc}{\widetilde{N}_c}
\newcommand{\tilq}{{\widetilde{q}\,}}
\newcommand{\tiLambda}{\widetilde{\Lambda}}

\def\noj#1,#2,{{\bf #1} (19#2)\ }
\def\jou#1,#2,#3,{{\sl #1\/ }{\bf #2} (19#3)\ }
\def\ann#1,#2,{{\sl Ann.\ Physics\/ }{\bf #1} (19#2)\ }
\def\cmp#1,#2,{{\sl Comm.\ Math.\ Phys.\/ }{\bf #1} (19#2)\ }
\def\ma#1,#2,{{\sl Math.\ Ann.\/ }{\bf #1} (19#2)\ }
\def\ng#1,#2,{{\sl Nagoya.\ Math.\ J.\/ }{\bf #1} (19#2)\ }
\def\jd#1,#2,{{\sl J.\ Diff.\ Geom.\/ }{\bf #1} (19#2)\ }
\def\invm#1,#2,{{\sl Invent.\ Math.\/ }{\bf #1} (19#2)\ }
\def\cq#1,#2,{{\sl Class.\ Quantum Grav.\/ }{\bf #1} (19#2)\ }
\def\cqg#1,#2,{{\sl Class.\ Quantum Grav.\/ }{\bf #1} (19#2)\ }
\def\ijmp#1,#2,{{\sl Int.\ J.\ Mod.\ Phys.\/ }{\bf A#1} (19#2)\ }
\def\jmphy#1,#2,{{\sl J.\ Geom.\ Phys.\/ }{\bf #1} (19#2)\ }
\def\jams#1,#2,{{\sl J.\ Amer.\ Math.\ Soc.\/ }{\bf #1} (19#2)\ }
\def\grg#1,#2,{{\sl Gen.\ Rel.\ Grav.\/ }{\bf #1} (19#2)\ }
\def\mpl#1,#2,{{\sl Mod.\ Phys.\ Lett.\/ }{\bf A#1} (19#2)\ }
\def\nc#1,#2,{{\sl Nuovo Cim.\/ }{\bf #1} (19#2)\ }
\def\np#1,#2,{{\sl Nucl.\ Phys.\/ }{\bf B#1} (19#2)\ }
\def\pl#1,#2,{{\sl Phys.\ Lett.\/ }{\bf #1B} (19#2)\ }
\def\pla#1,#2,{{\sl Phys.\ Lett.\/ }{\bf #1A} (19#2)\ }
\def\pr#1,#2,{{\sl Phys.\ Rev.\/ }{\bf #1} (19#2)\ }
\def\prd#1,#2,{{\sl Phys.\ Rev.\/ }{\bf D#1} (19#2)\ }
\def\prl#1,#2,{{\sl Phys.\ Rev.\ Lett.\/ }{\bf #1} (19#2)\ }
\def\prp#1,#2,{{\sl Phys.\ Rept.\/ }{\bf #1C} (19#2)\ }
\def\ptp#1,#2,{{\sl Prog.\ Theor.\ Phys.\/ }{\bf #1} (19#2)\ }
\def\ptpsup#1,#2,{{\sl Prog.\ Theor.\ Phys.\/ Suppl.\/ }{\bf #1} (19#2)\ }
\def\rmp#1,#2,{{\sl Rev.\ Mod.\ Phys.\/ }{\bf #1} (19#2)\ }
\def\yadfiz#1,#2,#3[#4,#5]{{\sl Yad.\ Fiz.\/ }{\bf #1} (19#2) #3%
\ [{\sl Sov.\ J.\ Nucl.\ Phys.\/ }{\bf #4} (19#2) #5]}
\def\zh#1,#2,#3[#4,#5]{{\sl Zh.\ Exp.\ Theor.\ Fiz.\/ }{\bf #1} (19#2) #3%
\ [{\sl Sov.\ Phys.\ JETP\/ }{\bf #4} (19#2) #5]}

\def\beq{\begin{equation}}
\def\eeq{\end{equation}}
\def\beqar{\begin{eqnarray}}
\def\eeqar{\end{eqnarray}}

\newcommand{\be}{\begin{equation}}
\newcommand{\ee}{\end{equation}}
\newcommand{\bea}{\begin{eqnarray}}
\newcommand{\eea}{\end{eqnarray}}

\def\nfrac#1#2{{\displaystyle{\vphantom1\smash{\lower.5ex\hbox{\small$#1$}}%
        \over\vphantom1\smash{\raise.25ex\hbox{\small$#2$}}}}}

\def\n#1{\mskip-#1mu}

\def\to{\rightarrow}

\def\lae{\mathrel{\mathop{\smash{\lower .5 ex \hbox{$\stackrel<\sim$}}}}}
\def\lae{\mathrel{\mathop{\smash{\lower .5 ex \hbox{$\stackrel>\sim$}}}}}

\def\Tr{{\rm Tr}}
\def\l:{\mathopen{:}\,}
\def\r:{\,\mathclose{:}}

\catcode`\@=11
\def\theequation{\arabic{equation}}
\catcode`\@=12

\nblack
\bcites

\nblack

\catcode`\@=11
\def\theequation{\thesection.\arabic{equation}}
\@addtoreset{equation}{section}
\@addtoreset{footnote}{section}
\@addtoreset{footnote}{subsection}
\catcode`\@=12

\newcommand{\beqn}{\begin{equation}}
\newcommand{\eeqn}{\end{equation}}
\newcommand{\beqnarray}{\begin{eqnarray}}
\newcommand{\eeqnarray}{\end{eqnarray}}
\newcommand {\bear} [1] {\begin {array} {#1}}
\newcommand {\ear} {\end {array}}

\newcommand{\CP}{{\bf C}{\rm P}}

\newcommand {\beqarn} {\begin{eqnarray*}}
\newcommand {\eeqarn} {\end{eqnarray*}}

\begin{document}

\begin{titlepage}

\begin{center}
\today
\hfill LBNL-41227, UCB-PTH-97/71\\
\hfill                  hep-th/9805142

\vskip 1.5 cm
{\large \bf Branes and Electric-Magnetic Duality in
Supersymmetric QCD}
\vskip 1 cm 
{Kentaro Hori}\\
\vskip 0.5cm
{\sl Department of Physics,
University of California at Berkeley\\
366 Le\thinspace Conte Hall, Berkeley, CA 94720-7300, U.S.A.\\
and\\
Theoretical Physics Group, Mail Stop 50A--5101\\
Ernest Orlando Lawrence Berkeley National Laboratory\\
Berkeley, CA 94720, U.S.A.\\}

\end{center}

\vskip 0.5 cm
\begin{abstract}
\noindent
We study the Type IIA limit of the \MT fivebrane configuration
corresponding to $N=1$ supersymmetric QCD with massless quarks.
We identify the effective gauge coupling constant
that fits with
Novikov-Shifman-Veinshtein-Zakharov exact beta function.
We find two different Type IIA limits that correspond to
the electric and magnetic descriptions of SQCD, as observed
in the massive case by Schmaltz and Sundrum.
The analysis is extended to the case of
symplectic and orthogonal gauge groups.
In any of the cases considered in this paper,
the electric and magnetic configurations are smoothly interpolated
via $M$ theory.
This is in sharp contrast with the proposed derivation of
$N=1$ duality within the
weakly coupled Type IIA string theory
where a singularity is inevitable unless
one turns on a parameter that takes the theory away from
an interesting point.

\end{abstract}

\end{titlepage}

\section{Introduction}

Supersymmetric gauge theories in various dimensions
have been studied
by realizing them on the worldvolume of branes and using superstring
duality \cite{GK}.
In particular,
the duality between Type IIA string theory
and eleven-dimensional supergravity can be used to
learn about four-dimensional gauge theories
\cite{W2,HOO,W1,BIKSY}.
Type IIA brane configuration, on which a gauge theory is realized,
is in general singular and hard to analyze, but
it becomes smooth and tractable in the eleven-dimensional
supergravity limit where the Type IIA string
coupling constant $\gst$ becomes large.

The worldvolume theory is coupled to various string excitation modes
and depends on parameters such as $\gst$
which have no counterpart in ordinary gauge theory.
In particular the theory becomes quantitatively different
as we send $\gst$ large, as explicitly shown in \cite{DHOO}.
However, some quantities (mainly, BPS or holomorphic objects)
are independent of such extra parameters
and can be computed by going to a region of the parameter space
where the supergravity approximation becomes valid.
In addition, as far as there is no singularity in going to such a
region, it is likely that qualitative features are unchanged.
One remarkable prediction based on this assumption
is that the QCD strings
can end on domain walls in $N=1$ supersymmetric Yang-Mills theory
\cite{W1}.

In this paper, we show that the two Type IIA configurations of
\cite{EGK} realizing the massless $N=1$ SQCD and the dual magnetic
theory \cite{Seiberg} are smoothly interpolated
by a family of fivebrane configurations in $M$ theory.
The smoothness of the interpolation strongly suggests that
the universality class of the theory remains the same through the
process.

An interpolation of the two configuration
was earlier considered within the weakly coupled Type IIA string
theory in \cite{EGK} and in many subsequent papers cited in
 \cite{GK} (see also \cite{OV} for a differnt but closely related
approach).
However, one must turn on a parameter corresponding to the
Fayet-Iliopoulos parameter of the $U(1)$ factor of the gauge
group in order to avoid a singularity in the process of
brane move.
It is doubtful whether there is a $U(1)$ factor
in the gauge group,
but even if we admit the $U(1)$ factor,
turning on the FI parameter takes the theory away from
an interesting point of the moduli space, and the brane move
simply leads to an equivalence of two free or empty theories.
One may turn on and off the FI parameter in the course of the
brane move, but a similar manipulation, such as turning
on and off the mass parameter, will send the theory to the
one with less flavors which is definitely different from the
original one. 
Furthermore, in the case of symplectic or orthogonal gauge groups,
there is no room for such manipulation \cite{EJS,EGKRS}
and the singularity
seems unavoidable.

What is shown in this paper can be considered as the \MT
resolution of such a singular brane move in Type IIA string theory.
There is no singularity in the process of interpolation 
even in the case of symplectic and orthogonal gauge groups,
and there is no need to change the physical parameters.
We also show that the Type IIA brane move with
mass parameters turned on and off, as mentioned
above, involves a singularity in $M$ theory side
and the change of the universality class can be explained.

Interpolating the electric and magnetic configurations
via \MT was first discussed in \cite{SS} where
$N=1$ SQCD with massive quarks was considered.
Extension to other groups and interactions, again with mass terms for
quarks, has also appeared in subsequent works
\cite{CS,CSST,Evans,Sugi}.
Turning on mass parameter takes the theory away from an interesting
point, but the basic idea can be extended to the massless case
and will be used in this paper.

This paper is organized as follows.

In section 2, we identify the holomorphic effective gauge coupling at
an energy higher than the dynamical scale using the
fivebrane configuration whose complex structure is
determined in \cite{HOO}.
We show that the wavefunction normalization factor enters
in a way compatible with
Novikov-Shifman-Veinshtein-Zakharov (NSVZ) exact beta function
\cite{NSVZ}.

In section 3, we consider the Type IIA limit of the \MT
fivebrane configuration and show that there are 
two limits that correspond to the electric
and magnetic theories of \cite{Seiberg}.
We discuss what kind of brane move does not change the universality
class of the theory.
Some technical details relevant in this section is implicitly
given in \cite{HOO} section 5.4. and are presented
more explicitly in \cite{NOYY,Y,NOY}.

In section 4, we extend the analysis to the
case of symplectic gauge group.
We show that Type IIA configurations realizing
the electric and magnetic theories
are smoothly interpolated through a family of
fivebrane configurations.
We also show that the
fivebrane configuration for a dynamical supersymmetry breaking model
proposed in \cite{DSB} has the correct Type IIA limit.

In section 5, we consider the orthogonal groups.
We again show that Type IIA configurations realizing
the electric and magnetic theories
are smoothly interpolated through a family of
fivebrane configurations.
We uncover a problem in describing exact results in terms of branes
for the case $N_f=N_c-2$ where the theory flows generically
to a Maxwell theory.

In an appendix, we correct an error in a field theory argument
of \cite{HOO},
and show that the brane prediction on the supersymmetric ground
states was correct.

\section{Branes and Renormalization Group}

In the study of Type IIA brane configurations corresponding to $N=2$
supersymmetric gauge theories in four dimensions \cite{W2},
the bending of NS5-branes caused by D4-branes
ending on them was interpreted as reflecting the running of the gauge
coupling constant.
In this section, we extend this interpretation
to the case of $N=1$ supersymmetric QCD, by considering
a suitable Type IIA limit of the configuration of
\MT fivebrane constructed in \cite{HOO}.
We will see that the wavefunction normalization factor
enters correctly in the formula for
the effective gauge coupling constant,
which is relevant for the derivation of
NSVZ beta function of \cite{NSVZ}.

\subsection{A Simple Example}

We start with reviewing the \MT fivebrane configuration for
$N=2$ $SU(N_c)$ super-Yang-Mills theory.
The configuration is in the flat eleven-dimensional spacetime
where the eleventh direction is compactified on a circle of radius
$R$. We introduce the time and space cooredinates
$x^0,x^1,\ldots,x^{10}$ with respect to which the flat metric
is expressed as
\beq
\dd s^2=-(\dd x^0)^2+(\dd x^1)^2+\cdots+(\dd x^9)^2+R^2(\dd x^{10})^2,
\eeq
where $x^{10}$ parametrize the compact eleven-th direction,
$x^{10}\equiv x^{10}+2\pi$.
The worldvolume of the fivebrane spans the four-dimensional space-time
in the $x^{0,1,2,3}$ directions and wraps on a two-dimensional surface
embedded in the four-manifold $\R^3\times S^1$
at $x^7=x^8=x^9=0$ parametrized by $x^{4,5,6,10}$.
We give a complex structure to
$\R^3\times S^1$ by a choice of complex coordinates
\beqa
&&v=\elst^{-2}(x^4+ix^5),
\label{defv}\\
&&t=\exp\Bigl(-{x^6\over R}-ix^{10}\Bigr).
\eeqa
In (\ref{defv}) we introduced the prefactor
$\elst^{-2}$, the tension of Type IIA strings,
so that $|{\sl \Delta}v|$ measures the mass of
an open string stretched between an interval
${\sl \Delta}v$.

The supersymmetric
configuration for $N=2$ $SU(N_c)$ super-Yang-Mills
theory is given by the fivebrane wrapped on the Riemann surface
in the $t$-$v$ space described by
\beq
t+t^{-1}={1\over \Lambda^{N_c}}\prod_{a=1}^{N_c}(v-\phi_a),
\label{N=2YM}
\eeq
where $\Lambda$ is a parameter that will be identified with the
dynamical scale of the super-Yang-Mills theory.
The parameters $\phi_a$ are interpreted as the eigenvalues of
the vev of the adjoint chiral superfield. 

\begin{figure}[htb]
\begin{center}
\epsfxsize=2.1in\leavevmode\epsfbox{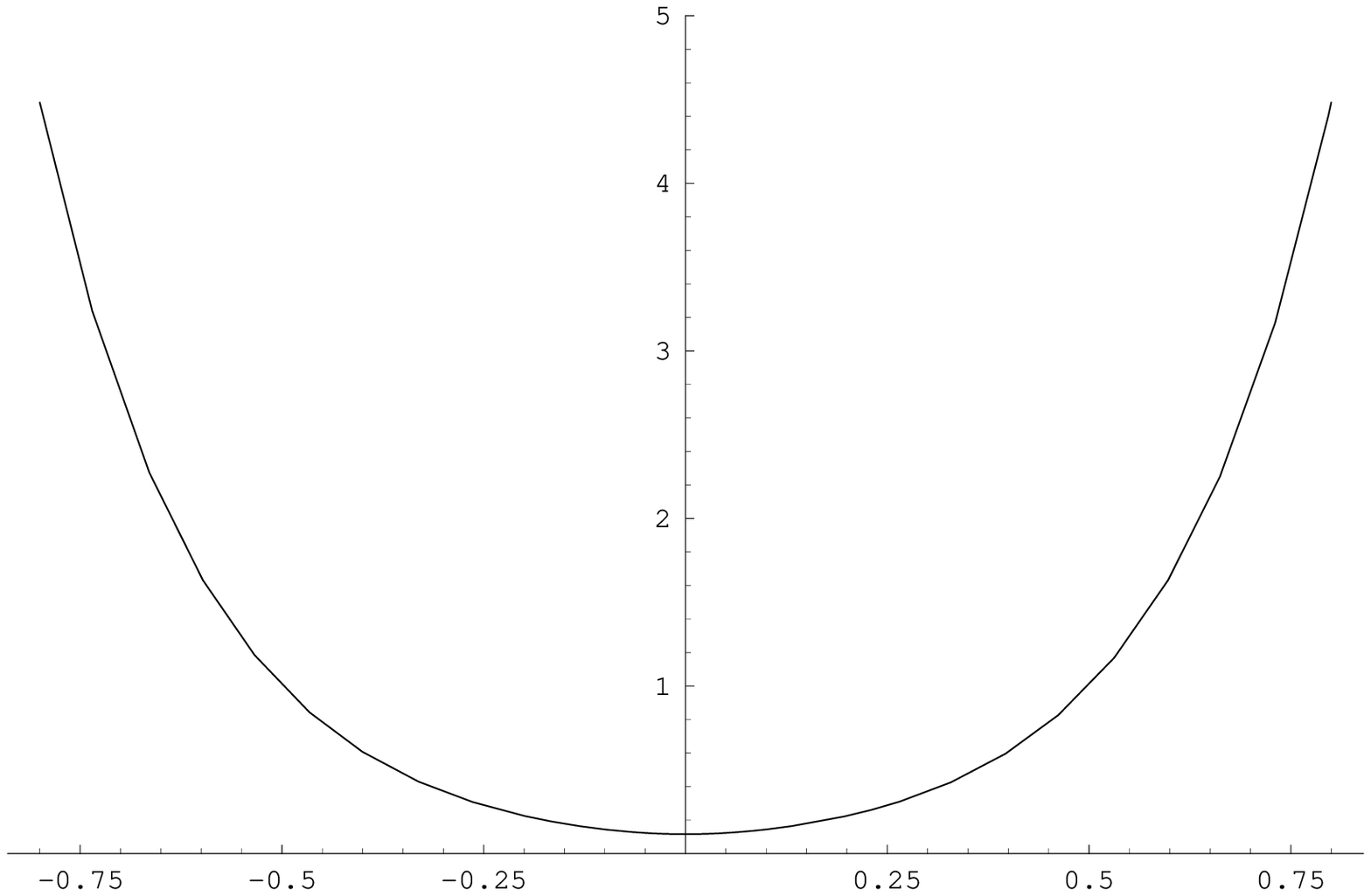}
\qquad
\epsfxsize=2.1in\leavevmode\epsfbox{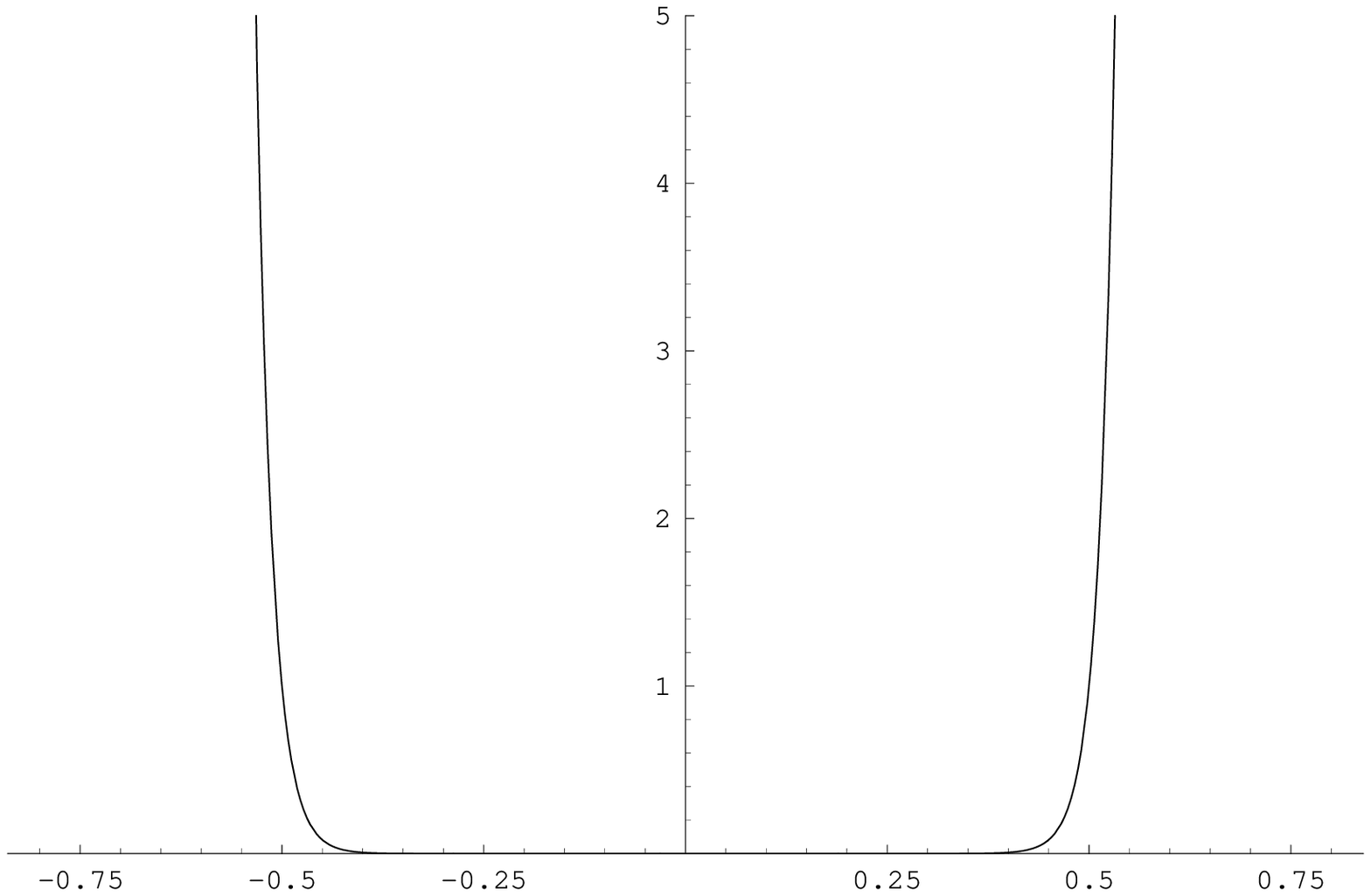}
\end{center}
\caption{Graphs of
$|v|=\sqrt{\e^{-{0.5\over R}}(\e^{x^6\over R}+\e^{-{x^6\over R}})}$ for
$R=0.1$ (left) and $R=0.01$ (right)}
\label{01}
\end{figure}

For the values of $R$ and $\Lambda$ small
compared to some length $L$ and energy $E$, the Riemann surface 
(\ref{N=2YM}) looks like two flat sheets connected by thin necks,
with respect to the scales set by $L$ and $E$
in the $x^6$ and $v$ directions.
The distiction between the regions of the flat sheets
and thin necks becomes sharper as $L/R$ and $E/\Lambda$ becomes
larger.
If we send these quantities very large by
keepng the combination $\e^{L/2R}(\Lambda/E)^{N_c}$
to be finite, say $1$, the sheet regions and the neck region
are separated at $x^6=\pm L/2$.
See Figure \ref{01} in which we plot $|v|$
as a function of $x^6$ (for a fixed $x^{10}$)
in the case of $N_c=2$ where we set $L=E=1$
and we put $\phi_a=0$.

In the weakly coupled Type IIA limit $R\ll\elel$, 
the flat sheets and thin necks correspond to the NS5-branes and
D4-branes respectively.
Then, $L$ is interpreted as the length of the D4-branes stretched
in the $x^6$ direction between the two NS5-branes.
Since $1/R=1/(\gst\elst)$ is the inverse
square of the gauge coupling constant of the
$4+1$-dimensional theory of the D4-brane worldvolume,
$L/R$ is identified as
the inverse square of the gauge coupling constant of the
four-dimensional theory obtained by
compactification on the interval of length $L$:
\beq
{L\over R}={1\over g^2}.
\label{N=2coupling}
\eeq
As we change $E$ with $\Lambda$ being fixed,
the ratio $L/R$ changes according to
\beq
\e^{-L/R}=\left({\Lambda\over E}\right)^{2N_c}\,.
\label{N=2match}
\eeq
In \cite{W2}, this was interpreted as reflecting the running
of the gauge coupling constant.

Actually, $g^2$ in (\ref{N=2coupling}) can be indentified
as the effective gauge coupling constant at the energy scale $E$
of the asymptotic free $SU(N_c)$ Yang-Mills theory. In particular,
(\ref{N=2match}) can be used to identify
the parameter $\Lambda$ as the dynamical scale of the theory.
This follows from the following reasoning.
Consider for instance two D4-branes sepatated by a distance
${\sl \Delta}\ell$ which is smaller than the string length $\elst$.
At energies below the string scale $M_{\it st}=1/\elst$,
the theory on
the worldvolume is the $U(2)$ super-Yang-Mills theory.
The non-abelian gauge symmetry is unbroken at energies above
the W-bosin mass $\elst^{-2}{\sl\Delta}\ell$, but is broken to the
abelian subgroup below this scale.
In other words, at a fixed energy
scale $E$ below $M_{\it st}$,
the $SU(2)$ gauge symmetry is unbroken if the separation is
small enough ${\sl\Delta}\ell<\elst^2E$.
Let us go back to our configuration.
At large values of $|x^6|$ such that $v$ is much larger than
$\Lambda$ and $\phi_a$'s, the curve is roughly described by
$v=\Lambda \e^{|x^6|/N_cR\pm ix^{10}/N_c}$.
In the interval $-L/2<x^6<L/2$, the configuration looks as the
$N_c$ D4-branes which are nearly on top of each other
compared to the length scale $\elst^2E$,
since the value of $v$ is $E$ at $x^6=\pm L/2$ and exponentially
small as $|x^6|$ descreases.
Thus, the $SU(N_c)$ gauge symmetry is unbroken at the energy
$E$ in this interval.
Outside the interval,
the separation of the D4-branes is much greater than $\elst^2E$
and the W-bosons decouple at an energy above $E$.
Or more precisely, we see that the picture of aligned D4-branes
is actually incorrect in this region and there is no obvious
candidate for the W-bosons.
In any case, we do not have any non-trivial gauge dynamics at the
energy $E$ from the region outside the interval $-L/2<x^6<L/2$.
Therefore we can consider $g$ given by (\ref{N=2coupling}) as
the gauge coupling constant set at the energy $E$
as far as $E<1/L$, and hence we can identify the parameter $\Lambda$
as the dynamical scale of the four-dimensional gauge theory.

The holomorphic description of the fivebrane
is of course valid in the
region of the parameter space where all the characteristic lengths
of the brane and space-time are much larger than the eleven-dimensional
Planck length, in particular
$R\gg \elel\gg\elst$ and $\Lambda\gg \elst^{-2}\elel\gg
M_{\it st}$.
Since this is not the region where the gauge dynamics is dominant,
one may wonder whether the measurement of the gauge
coupling based on the holomorphic description is valid.
The point is
that we can scale up the lengths $R,L$ and energies
$\Lambda,E$ to the region where holomorphic description is valid
while keeping the ratio $L/R$ and $\Lambda/E$ fixed.
Then, the gauge coupling constant
measured as (\ref{N=2coupling})-(\ref{N=2match})
is invariant under the rescaling, and therefore it can be considered
to be valid even when we scale down the parameters
back to the gauge theory region.

\subsection{$N=1$ Supersymmetric QCD}


We next consider a configuration of the fivebrane corresponding to
$N=1$ SQCD with gauge group $SU(N_c)$ and $N_f\geq N_c$ flavors.
The complex structure of the configuration was determined in
\cite{HOO} as we will revisit shortly.

The fivebrane configuration is in the eleven-dimensional space-time
which is a product of the flat seven-dimensional space-time
(in the $x^{0,1,2,3,7,8,9}$ directions)
and the Taub-NUT space of $A_{N_f-1}$ type
(in the $x^{4,5,6,10}$ directions).
Thus, the metric of the eleven-dimensional space-time
is given by
\beq
\dd s^2=-(\dd x^0)^2+(\dd x^1)^2+(\dd x^2)^2 +(\dd x^3)^2
+\dd s_{\rm TN}^2+(\dd x^7)^2+(\dd x^8)^2+(\dd x^9)^2
\eeq
where $\dd s_{\rm TN}^2$ is the Taub-NUT metric which will be given
below.
The fivebrane spans the $x^{0,1,2,3}$ directions and wraps
on a Riemann surface in the transversal directions.

The Taub-NUT geometry in \MT corresponds to D6-branes in Type IIA string
theory which can be used to realize the quark multiplets.
If the D6 branes are at
$\vec{x}=(x^4,x^5,x^6)=\vec{x}_i$, $i=1,\ldots, N_f$,
the Taub-NUT metric is given by
\beq
\dd s_{\rm TN}^2\,=\,
U\,\dd \vec{x}^2
+R^2\,U^{-1}\left(\dd x^{10}+\vec{\omega}\cdot\dd\vec{x}\right)^2,
\label{TN}
\eeq
where
\beq
U\,=\,1+\sum_{i=1}^{N_f}\frac{R}{2|\vec{x}-\vec{x}_i|}\,,~~~
{\rm curl}\,\vec{\omega}={\rm grad}
\sum_{i=1}^{N_f}\frac{1}{|\vec{x}-\vec{x}_i|}\,.
\label{Uom}
\eeq
Since $x^{10}$ is a periodic coordinete of period $2\pi$,
the Taub-NUT space can be considered as a circle bundle over the
$\vec{x}$ space where the radius $R/\sqrt{U}$ shrinks at the position
of the sixbranes.
At large $|\vec{x}|$, the asymptotic radius is $R$ and
the equation (\ref{Uom}) shows that the circle bundle is the monopole
bundle of magnetic charge $N_f$.
In what follows, we place the sixbranes at $N_f$ points
in the $x^6$-axis $x_i^4=x_i^5=0$
so that the configuration we will consider corresponds
to massless SQCD.

\subsection*{\sl Holomorphic Description of the Configuration}

\newcommand{\const}{{\rm const}}

In \cite{HOO} we have given a holomorphic description of the
fivebrane configuration which we now review.

We consider the space transverse to the four
dimensions $x^{0,1,2,3}$ to be the product of a real line
(the $x^7$ direction)
and a three-dimensional complex manifold
(product of the Taub-NUT space and the $x^{8,9}$ plane).
We select one of the complex structures of the Taub-NUT space
such that
\beq
v=\const\times (x^4+ix^5)
\label{defvv}
\eeq
is a complex coordinate. 
We also provide the $x^{8,9}$ plane with a complex coordinate
\beq
w=\const\times (x^8+ix^9).
\label{defw}
\eeq
Under the choice (\ref{defvv}) of the complex structure,
other complex coordinates of the Taub-NUT space can be given by
\beqa
&&y=c\times\e^{-(x^6/R+ix^{10})}\prod_{i=1}^{N_f}
\sqrt{|\vec{x}-\vec{x}_i|-(x^6-x^6_i)}\,,
\label{defy}\\
&&x=c^{-1}\times\left({v\over |x^4+ix^5|}\right)^{N_f}
\e^{x^6/R+ix^{10}}\prod_{i=1}^{N_f}
\sqrt{|\vec{x}-\vec{x}_i|+(x^6-x^6_i)}\,,
\label{defx}
\eeqa
where $c$ is a constant.
It is a matter of convention that the proportionality
constant in (\ref{defx})
is chosen to be inverse to $c$ in (\ref{defy}).
The complex coordinates $y$, $x$ and $v$ are
related by
\beq
yx=v^{N_f}.
\label{Asing}
\eeq
It appears from this that the Taub-NUT space has a singularity
at $y=x=v=0$ which is indeed the singularity of $A_{N_f-1}$ type
when the sixbranes are on top of each other.
However, when the D6-branes are separated in the $x^6$ direction,
the singular point $y=x=v=0$ is blown up to $N_f-1$ $\CP^1$ cycles
and
the singularity is resolved as we can see by introducing an appropriate
coordinate system around each of the D6-branes (see the next section).

The fivebrane is wrapping a holomorphic curve
consisting of two infinite components
and several $\CP^1$ components.
The infinite components are described by
\beq
C^{\prime}~
\left\{
\begin{array}{l}
y=\prod_{i=1}^{N_c}(w-M_i)\\
v=0
\end{array}
\right.
~~~~~
C~\left\{
\begin{array}{l}
x=\Lambda^{-(3N_c-N_f)}v^{N_c}\\
w=0,
\end{array}
\right.
\label{HOOconf}
\eeq
while the $\CP^1$ components at $y=x=v=0$ will be described
in the next section.

This was obtained in \cite{HOO} by rotation and deformation of
the configuration corresponding to the $N=2$ SQCD.
($\widetilde{y}$ in \cite{HOO} is denoted simply by $y$ here and
the normalization of $x$ is also changed.)
The parameter $\Lambda$ in the above expression was identified in \cite{HOO}
as the dynamical scale of the $N=1$ SQCD
by imposing the renormalization group matching
with the high energy $N=2$ theory. In this paper, we will
see this directly by identifying the gauge coupling constant.
The parameters $M_i$ are interpreted as the eigenvalues of the
vacuum expectation value of the
meson matrix $M$.
The fact that the number of $M_i$'s
is $N_c$ reflects the
classical (= quantum) constraint ${\rm rank}M\leq N_c$ on the meson
matrix.
In addition, by identifying the parameters corresponding to the baryon
vevs, we can derive in the brane description
the quantum modified constraint for the theory with $N_f=N_c$
as well as the classical (= quantum) constraint for higher
$N_f>N_c$ concerning the relation between meson and baryon vevs
\cite{HOO}.\footnote{In \cite{HOO}, there was
a gap between the brane and field theory analysis
on the quantum constraint for the system with a heavy adjoint chiral
multiplet. In Appendix, we will show that this gap is because of
our error in the field theory analysis and that the brane
prediction is correct.}

\subsection*{\sl Extra Parameters}

The description of the holomorphic curve given above
does not completely specify the fivebrane configuration
in the eleven-dimensional space-time
since we have left undetermined the proportionality constants in 
(\ref{defvv})-(\ref{defy}).

These constants are constrained
if we require that the brane must correctly reproduce
the superpotential of the gauge theory.
In \cite{W1} the superpotential of the worldvolume theory
was defined as a certain integral of the holomorphic three-form
of the space-time Calabi-Yau three-fold in which the curve is embedded.
\footnote{The holomorphic three-form of a Calabi-Yau three-fold
is a nowhere vanishing holomorphic three-form such that its wedge product
with its complex conjugate is the same as the cube of the K\"ahler form.}
In the present case, the three-fold is
the product of the Taub-NUT space and the $x^{8,9}$ plane
with the complex structure chosen above, and the three-form
is $\dd(x^4+ix^5)\wedge\dd(x^8+ix^9)\wedge\dd(x^6+iRx^{10})$
at large $|x^{4,5,6}|$.
As has been shown in appendix A of \cite{DHOO} in
the example of $N=1$ super-Yang-Mills theory,
in order for the brane computation to reproduce the field theory result,
the holomorphic three-form
multiplied by the fivebrane tension $\elel^{-6}$ 
must be given by
\beq
\Omega=\dd v\wedge\dd w\wedge{\dd y\over y}\,.
\label{Om}
\eeq
This requires the product of the constants in
(\ref{defvv}) and (\ref{defw}) to be equal to $R\elel^{-6}$.
Then, the coordinates $v$ and $w$ can be expressed as
\beq
v=\elst^{-2} Z(x^4+ix^5),~~~
w=\elel^{-3} Z^{-1}(x^8+ix^9),
\label{defvw}
\eeq
where $Z$ is a constant.

Actually, there is a redundancy in the parametrization.
Obviously, 
a phase shift of the constant $c$ in (\ref{defy})-(\ref{defx})
can be compensated by a shift of the origin of $x^{10}$.
Also, a change of the magnitude of $c$
can be absorbed by an overall
shift of $x_i^6$'s, and only the combination
$c\,\e^{-\sum_ix^6_i/R}$ characterizing the relative separation
of the fivebrane system from the sixbranes in the $x^6$ direction
has an invariant meaning.
Furthermore,
the configuration is invariant under the change of parameters
as
\beqa
&&Z\to\zeta^{-1}Z,~~~ c\to\zeta^{N_c}c,\label{resc0}\\
&&M_i\to \zeta M_i,
\label{resc}\\
&&\Lambda^{3N_c-N_f}\to\zeta^{N_f}\Lambda^{3N_c-N_f}
\label{KSanom}
\eeqa
as it can be compensated by the coordinate transformation $v\to\zeta^{-1}v$,
$w\to \zeta w$, $y\to \zeta^{N_c}y$ and $x\to \zeta^{-N_c-N_f}x$.
This transformation can be identified with the
action of the complexified axial $U(1)$ flavor symmetry group
which is anomalously broken to $\Z_{N_f}$.
The radial part is the group of rescalings of
the quark field as can be seen in (\ref{resc}),
and the transformation of $\Lambda$ in (\ref{KSanom}) reflects
its anomaly \cite{KS} (see also \cite{AM})
as we will see after identifying the gauge coupling constant.

We conclude that the configuration depends on two real
parameters, $Z^{N_c}c\,\e^{-\sum_i x^6_i/R}$ and $R$,
as well as the separation
$x^6_i-x^6_{i+1}$ of the D6-branes, in addition to the
parameter and observables of the four-dimensional gauge theory
($\Lambda$, $M_i$'s and the location and the chiral two-form
of the $\CP^1$ components)

\subsection*{\sl The Effective Gauge Coupling Constant}

As in the case of $N=2$ super-Yang-Mills theory,
we want to distinguish the region of the fivebrane
where the gauge dynamics is relevant at an energy $E$
below the string scale $1/\elst$,
by setting an appropriate length scale corresponding to $E$.
In the present case, the fivebrane is extending in two
directions, $v$ and $w$, and we must set a scale
for each of them.
We propose that these scales, $v(E)$ and $w(E)$, are related by
$v(E)w(E)=E^3$.
This is based on the fact that
the holomorphic three-form $\Omega=\dd v\wedge\dd w\wedge \dd y/y$
should transform as $\Omega\to \lambda^3\Omega$
under a change of the energy scale $E\to \lambda E$
since the superpotential is given by an integral of $\Omega$
and has mass dimension three.

Let us first determine the scale in the $v$ direction.
Note that only the component $C$ is extending in this direction.
Since this part of the fivebrane is the same as a part of the $N=2$
configuration, as far as the effect of the other part of the fivebrane
is negligible, we can say that
the gauge dynamics is relevant in the region where the
configuration looks as nearly coincident D4-branes compared to
the length scale $\elst^2 E$ just as in the $N=2$ theory.
Since the coordinates $x^{4,5}$ and $v$ are related by (\ref{defvw}),
this region is given by $|v|<ZE$.
Note that the value of $y$ at the boundary $v=ZE$ is
\beq
y(E)=\Lambda^{3N_c-N_f}(ZE)^{N_f-N_c}.
\label{yE}
\eeq
Since the scale in the $v$-direction is set by $v(E)= ZE$,
the scale in the $w$ direction is set by $w(E)= Z^{-1}E^2$.
This corresponds to a length scale $\elel^3Zw\sim\elst^2 RE^2$ which
is much shorter than $\elst^2E$ in the weakly coupled Type IIA
limit $E< 1/\elst\ll 1/R$.
Actually, we can provide the following interpretation
of this new scale $w(E)\propto E^2$.
The $N_c$ $w$-values of the component $C^{\prime}$
for a fixed $y$ can be interpreted as the non-zero
eigenvalues of the meson matrix $\tilQ Q$ at $y$,
as these values at $y=0$ are interpreted as their
vacuum expectation values.
Since non-zero values of $Q,\tilQ$
give a mass of order $\sqrt{\tilQ Q}$ to the gluons,
the gauge symmetry is unbroken at the energy $E$ in the
region where $\sqrt{w}$ is smaller than a value of order $E$.
The precise value $Zw= E^2$ suggests that the quark kinetic term
is given by $Z(Q^{\dag}Q+\tilQ\tilQ^{\dag})$ and the
mass of the gauge boson is $\sqrt{Z\tilQ Q}$.
Thus, the gauge dynamics is
important at the energy $E$ only in the region $\sqrt{|Zw|}<E$, or
$|Zw|<E^2$,
where $|y|$ is smaller
than
\beq
y^{\prime}(E)=(Z^{-1}E^2)^{N_c}.
\label{ypE}
\eeq
To summarize, the $SU(N_c)$ gauge dynamics is relevant in the region
$|v|<ZE$ and $|w|<Z^{-1}E^2$,
and the value of $y$ at the boundaries are
$y(E)$ and $y^{\prime}(E)$ given by (\ref{yE}) and (\ref{ypE})
respectively.

We now want to
measure the effective gauge coupling constant $g(E)$ at the 
energy $E$. 
In the case of $N=2$ super-Yang-Mills theory,
we have seen that $1/g^2(E)$ is $1/R$ times the length in the
$x^6$-direction of the part of the fivebrane
in which the gauge dynamics is relevant at the energy $E$.
In other words,
$\e^{-1/g^2(E)}$ is given by the ratio of the values of
the coordinate $t=\e^{-x^6/R-ix^{10}}$
at the two boundaries of the relevant part.
The holomorphic coordinate analogous to $t$
is $y$ in the present case:
It is the only complex
coordinate that has an asymptotic behaviour
$\sim \e^{-x^6/R-ix^{10}}$ at large $x^6$
and does not vanish at a generic point in both of the
components $C$ and $C^{\prime}$.
Therefore,
we propose that $\e^{-1/g^2(E)}$ is likewise
given by the ratio of the values of $y$ at the boundaries of
the relevant region, i.e. $y(E)/y^{\prime}(E)$.

This may require some explanation.
The logarithm of the ratio of the two values of $y$
is in general not close to $1/R$ times the distance
of the boundaries.
This is not a problem since the definition of the gauge coupling
constant is likely to be corrected by the intersection with
the sixbranes. At present, we do not know the precise way to
measure the gauge coupling constant in such a situation.
Fortunately, there is one parameter $c$ corresponding
to the relative position of the sixbranes and the fivebrane
system in the $x^6$-direction, which has no counterpart
in the four-dimensional gauge theory.
It is expected that the change of $c$ is irrelevant
in gauge theory as far as the curve is smoothly deformed.
Especially, a change of $c$ does not affect the holomorphic
object, such as holomorphic gauge coupling constant.
We can tune $c$ so that all the sixbranes are sent far to the right
($x^6_i$ large).
Then, the sixbranes goes away from the region sandwiched by the
two boundaries. In the limit $x^6_i\to+\infty$
where the effect of the intersection
with the sixbranes is suppressed,
$|{\rm log}(y(E)/y^{\prime}(E))|$ approaches
$1/R$ times the actual distance of the two boundaries.

We thus obtain
\beq
\e^{-1/g^2(E)}
={y(E)\over y^{\prime}(E)}=
\left({\Lambda\over E}\right)^{3N_c-N_f}Z^{N_f}.
\label{holg}
\eeq
This is the same as the one-loop running of the gauge coupling constant
of the SQCD.
The appearance of the factor $Z^{N_f}$ is actually consistent with
the interpretation of $Z$ as the normalization factor
of the quark kinetic term $Z(Q^{\dag}Q+\tilQ\tilQ^{\dag})$
with respect to the fields $Q,\tilQ$ such that
the meson matrix $\tilQ Q$ is identified with the
parameters $M_i$.
In order to show this, we first note that the gauge coupling in
$N=1$ SQCD depends on the choice of variables for the quark multiplet.
Rescaling of the quark superfields $Q\to\sqrt{\zeta}Q$,
$\tilQ\to\sqrt{\zeta}\tilQ$ generates a Jacobian which shifts the inverse
square of the gauge coupling constant as $\e^{-1/g^2}
\to\zeta^{N_f}\e^{-1/g^2}$ \cite{KS,AM}.
The standard and unambiguous prescription to define the gauge coupling
constant is to choose the variable such that the quark kinetic term is
canonically normalized at a given energy $E$.
Since we have made no choice of variables when defining the coupling
constant $g(E)$ as (\ref{holg}), it is natural to consider it
as gauge coupling constant defined in the standard way.
By using the rescaling symmetry (\ref{resc0}) with $\zeta=Z$,
we can make $Z\to 1$ and this induces the change of parameters as
$M_i\to ZM_i$ and $\Lambda^{3N_c-N_f}
\to Z^{N_f}\Lambda^{3N_c-N_f}$ (\ref{resc})-(\ref{KSanom}).
With respect to the new variables $M^{\prime}_i=ZM_i$
and $\Lambda^{\prime 3N_c-N_f}=Z^{N_f}\Lambda^{3N_c-N_f}$,
the quark kinetic term is canonically normalized and
the gauge coupling constant $g(E)$ has the standard
expression $\e^{-1/g^2(E)}=(\Lambda^{\prime}/E)^{3N_c-N_f}$.
The gauge coupling constant $\hat{g}(E)$ defined with respect to
the old variables $M_i$ is given by $\e^{-1/\hat{g}^2(E)}
=(\Lambda/E)^{3N_c-N_f}$ and the relation to the standard one
(\ref{holg}) reflects the rescaling anomaly of \cite{KS}.

In supersymmetric field theory, the gauge coupling constant
receives essentially only the one-loop correction,
and (\ref{holg}) describes the {\it exact} renormalization group
flow provided $Z$ changes according to the wavefunction renormalization
$Z\to Z^{\prime}$ under a change of the energy scale $E\to E^{\prime}$
\cite{NSVZ} (see also \cite{AM}).
In the brane picture, $Z$ is a fixed parameter and hence it appears that
the brane does not capture the effect of the wavefunction renormalization.
However, the renormalization $Z\to Z^{\prime}$ of the kinetic term of
$Q$, $\tilQ$ suggests that the scale in the $w$ direction must change from
$Zw(E)=E^2$ to $Z^{\prime}w(E^{\prime})=E^{\prime 2}$.
Then, the scale in the $v$ direction changes as $v(E^{\prime})
=Z^{\prime}E^{\prime}$, and the gauge coupling constant at the energy
$E^{\prime}$ is given by $\e^{-1/g^2(E^{\prime})}
=(\Lambda/E^{\prime})^{3N_c-N_f}Z^{\prime N_f}$.
The flow $g(E)\to g(E^{\prime})$ is nothing but what
arizes from the exact beta function of \cite{NSVZ}.
(To be precise,
the beta function of \cite{NSVZ} is concerned with ``1PI''
coupling whereas what we have been considering is the
holomorphic, or ``Wilsonian'' coupling. The relation between them
is known \cite{SV}, and translates the flow described above to the
flow of ``1PI'' coupling of \cite{NSVZ}.)
Note that the length scale in the $v$-direction has changed as
$\elst^2 E\to \elst^2{Z^{\prime}\over Z}E^{\prime}$.
This may be interpreted as the effect on the component $C$ 
from other components of the fivebrane, which we can no longer neglect
as $E$ is reduced since the boundary becomes
closer to the other component
$C^{\prime}$.

\subsection*{\sl $N=2$ Broken to $N=1$}

In order to examine the above idea to identify  the
gauge coupling constant, we consider $N=2$ SQCD with $N=2$ broken
to $N=1$ by giving a mass $\mua$ to the adjoint chiral multiplet.
This model has been analyzed in \cite{HOO} in the brane picture
and it was shown that there are various barnches of the moduli space.

When we only look at high energies compared to the scales
determined by the vevs or the dynamical scale there is no difference
between the various branches, and therefore we
only look at the simplest one,
the baryonic branch. The curve of the baryonic branch consists of
several $\CP^1$ components and the two infinite components given by
\beq
C^{\prime}~
\left\{
\begin{array}{l}
y=v^{N_c}\\
v=\mua^{-1}w,
\end{array}
\right.
~~~~~
C~\left\{
\begin{array}{l}
y=\Lambda_{N=2}^{2N_c-N_f}v^{N_f-N_c}\\
w=0.
\end{array}
\right.
\label{finmu}
\eeq
At an energy $E$,
the scale in the $v$ and $w$ directions are
set by $v(E)=E$ and $w(E)=E^2$ (here we set $Z=1$ ignoring the wavefunction
renormalization for simplicity).

The component $C$ extends only in the $v$ direction, and the
gauge dynamics is relevant at $E$ in the region with $|v|<E$.
The value of $y$ at the boundary is
$y(E)=\Lambda_{N=2}^{2N_c-N_f}E^{N_f-N_c}$.

On the other hand, the component $C^{\prime}$ extends in the mixed direction
$w=\mua v$, and the region where the gauge dynamics
is relevant depends on the value of $E$ compared to $\mua$.
For the values of $v$ above $v(E)=E$
the gauge group is broken by the adjoint Higgs
field  whereas $w\sim E^2$ is the scale
where the gauge group is broken by the quark field.
Note that the value of $v$ at the quark Higgsing scale is
$v_{\rm H}(E)=\mua^{-1}E^2$ in the component $C^{\prime}$.\\
$\bullet$ When the energy $E$ is much larger than $\mua$,
the adjoint Higgsing scale is smaller $v(E)\ll v_{\rm H}(E)$
and hence the relevant region
is $|v|<|v(E)|$ with
the boundary value of $y$ being $y^{\prime}(E)=E^{N_c}$.
Then, the gauge coupling constant is given by
\beq
\e^{-1/g^2(E)}=
{y(E)\over y^{\prime}(E)}={\Lambda_{N=2}^{2N_c-N_f}\over E^{2N_c-N_f}},
\eeq
and runs in the same way as in the original $N=2$ SQCD.\\
$\bullet$
When the energy $E$ is much smaller than $\mua$,
quark Higgsing scale is smaller $v_{\rm H}(E)\ll v(E)$,
and hence the relevant region is $|v|<|v_{\rm H}(E)|$
with the boundary value of $y$ being $y^{\prime}(E)=(\mua^{-1}E^2)^{N_c}$.
Thus, the gauge coupling constant is given by
\beq
\e^{-1/g^2(E)}=
{y(E)\over y^{\prime}(E)}={\mua^{N_c}\Lambda_{N=2}^{2N_c-N_f}\over
E^{3N_c-N_f}},
\eeq
and runs in the same way as in $N=1$ SQCD with the dynamical scale
$\Lambda$ given by $\Lambda^{3N_c-N_f}=\mua^{N_c}\Lambda_{N=2}^{2N_c-N_f}$.

This agrees with the behaviour of the effective gauge coupling
in field theory.

\section{Electric-Magnetic Duality}

In this section, we consider the Type IIA limit of the configuration
for $N=1$ SQCD in more detail.
Following the idea of \cite{SS},
we will find two different limits
which coincide with the Type IIA configurations of \cite{EGK}
corresponding to the electric and magnetic description
of the theory.
The two limits are smoothly interpolated, and hence we may expect the
universality class of the theory to be unchanged through the process.

\subsection*{\sl Sixbranes and $\CP^1$ Cycles}

As mentioned in the previous section,
the description of the Taub-NUT space in terms of
the coordinates $y,x,v$ breaks down
near the locus of the sixbranes.
As far as the sixbranes are separated (in the $x^6$ direction in the
present case), the Taub-NUT space is smooth
and is described by introducing one coordinate system $(y_i,x_i)$ in a
neighborhood $U_i$ of each of the D6-branes ($i=1,\ldots,N_f$).
These coordinates are related to $y,x,v$ by
\beq
y=y_i^ix_i^{i-1},~~x=y_i^{N_f-i}x_i^{N_f+1-i},~
\mbox{and}~\,v=y_ix_i.
\label{iyx}
\eeq
and are related to each other in the overlapping region
$U_i\cap U_{i+1}$ by
\beq
x_iy_{i+1}=1~\,\mbox{and}~\,\,y_ix_i=y_{i+1}x_{i+1}.
\label{reliyx}
\eeq
There are $N_f-1$ $\CP^1$ cycles
$C_1,\ldots,C_{N_f-1}$.
The cycle $C_i$ is defined
as the locus of $y_i=0$ in $U_i$ and $x_{i+1}=0$ in $U_{i+1}$.
Or equivalently, by considering the Taub-NUT space as the circle bundle
over the $x^{4,5,6}$ space,
$C_i$ can also be defined as the fibres
over the straight segment in the $x^6$ axis
stretched between the $i$-th and the $(i+1)$-th
D6-branes.
In particular, $C_{i-1}$ and $C_i$ intersect
transversely at the position of the $i$-th D6-brane.
The volume of the cycle $C_i$ is $|x^6_i-x^6_{i+1}|R$.

In the weak coupling Type IIA limit $R\ll \elst$, 
the $\CP^1$ cycle $C_i$ looks like a thin cigar connecting the two
D6-branes.
A fivebrane wrapped on $C_i$
becomes a D4-brane stretched beteen the $i$-th
and $(i+1)$-th D6-branes.
Likewise, a membrane wrapped on $C_i$
becomes an open string stretched beteen these D6-branes
and has a mass of order $\elel^{-3}|x^6_i-x^6_{i+1}|R=
\elst^{-2}|x^6_i-x^6_{i+1}|$.
Such membranes can be identified as the W-bosons on the worldvolume
of the D6-branes. In the limit where all the D6-branes become on top
of each other, there is an enhanced $SU(N_f)$ gauge symmetry
on the D6-brane worldvolume which is a flavor symmetry 
from the point of view of the worldvolume theory on the fivebrane.
Since this is present already for $N=2$ configurations,
it corresponds to the diagonal subgroup of the chiral flavor
symmetry group of the SQCD.

\subsection*{\sl $\CP^1$ Components of the Fivebrane}

In addition
to the two infinite components $C$ and
$C^{\prime}$ of (\ref{HOOconf}),
the fivebrane
consists of several components wrapped on the $\CP^1$ cycles $C_i$.
As explained in \cite{HOO},
the number of $\CP^1$ components wrapped on $C_1,\ldots, C_{N_f-1}$
are
\beq
\underbrace{N_c,\ldots,N_c}_{N_f-N_c},N_c-1,\ldots,2,1.
\label{compo}
\eeq

One of the equations describing $C$,
\beq
x=\Lambda^{-(3N_c-N_f)}v^{N_f},
\eeq
actually describes the whole configuration
projected to the Taub-NUT space.
Let us look at this equation in
the $i$-th patch $U_i$, the neighborhood of the $i$-th D6-brane.
The equation reads
$y_i^{N_f-i}x_i^{N_f+1-i}=\Lambda^{-(3N_c-N_f)}y_i^{N_c}x_i^{N_c}$.
If $i$ is smaller than or equal to $N_f-N_c$,
the right hand side has lower powers of $y_i$ and $x_i$,
and the equation splits to $y_i^{N_c}=0$,
$x_i^{N_c}=0$ and
$y_i^{(N_f-N_c)-i}x_i^{(N_f-N_c+1)-i}=\Lambda^{-(3N_c-N_f)}$.
From the first and the second equations, we see that the fivebrane
wraps $N_c$ times on $C_i$ ($i\leq N_f-N_c$).
The solution of the last equation is the infinite component $C$.
If $i$ is larger than $N_f-N_c$, the left hand side has lower
powers, and the equation splits to
$y_i^{N_f-i}=0$, $x_i^{N_f+1-i}=0$, and
$y_i^{i-(N_f-N_c)}x_i^{i-(N_f-N_c+1)}=\Lambda^{3N_c-N_f}$.
We see that the fivebrane wraps $(N_f-i)$-times on $C_i$ ($i>N_f-N_c$)
from the first and the second equation.
The solution of the last equation is again the infinite component $C$.
The solution to $y_1^{N_c}=0$ in the first patch is the projection of
the other infinite component $C^{\prime}$.

\begin{figure}[htb]
\begin{center}
\epsfxsize=5.6in\leavevmode\epsfbox{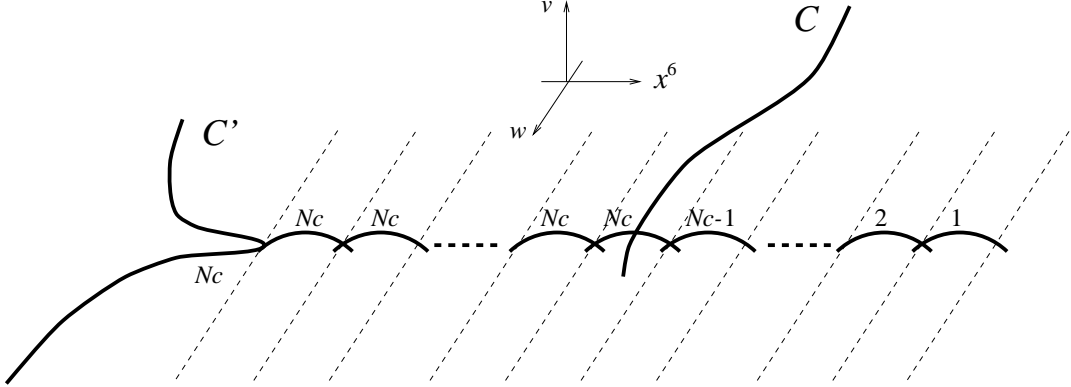}
\end{center}
\caption{The $M$ Theory Configuration of SQCD}
\label{Mconf}
\end{figure}

We depict in Figure \ref{Mconf}
the fivebrane configuration
at the origin of the moduli space (for the case with $N_f>N_c$).
The $\CP^1$ components can move
in the $x^{7,8,9}$ directions and the self-dual two form on them
can be turned on. The component $C^{\prime}$ can also
be deformed by turning on $M_i$ in (\ref{HOOconf}).
These deformations of the curve correspond to the flat directions
of the field theory.
\footnote{It is a matter of question whether the component $C^{\prime}$
can be translated in the $x^{7,8,9}$ directions and whether
its self-dual two form can be turned on ($x^{8,9}$
translation corresponds to turning on $\sum_iM_i$),
since it appears to cost an infinite amount of energy due to the
non-compactness of the component $C^{\prime}$.
The moduli space of the worldvolume theory have the same dimension
as the field theory moduli space
only when these degrees of freedom are considered as moduli
but not as parameters.}

\subsection{The Electric Limit}

Let us consider the limit
$\Lambda^{3N_c-N_f}\ll E^{3N_c-N_f}$
for a fixed energy $E$ and look at the fivebrane configuration
with respect to the scales set by $v\sim E$ and $w\sim E^2$.
(Here we put $Z=1$ by redefining $M_i$ and $\Lambda$ so that
the quark superfields have canonically normalized kinetic term
at the energy $E$.)
We also take the limit $R\ll L$ for a fixed length scale $L$
keeping $\e^{L/R}(\Lambda/E)^{3N_c-N_f}$ to be of order 1.
In the weakly coupled Type IIA limit $R\ll\elel$
with $E<1/\elst,1/L$,
we will see an intersecting Type IIA brane
configuration which corresponds to four-dimensional gauge theory.
The resulting configuration depends on how to tune the 
parameter $c$ in the limit.
We first consider $c$ to be fixed so that the Taub-NUT space remains
well-described by the coordinates $(y_i,x_i)$
except that the angular direction $y_i\to \e^{i\theta}y_i$,
$x_i\to \e^{-i\theta}x_i$ shrinks to zero size in the Type IIA limit.

As we have seen,
the infinite component $C$ is described in the $i$-th patch
by
\beqa
&&y_i^{(N_f-N_c)-i}x_i^{(N_f-N_c+1)-i}=\Lambda^{-(3N_c-N_f)}
,~~\mbox{for $1\leq i\leq N_f-N_c$}
\label{below}\\
&&y_i^{i-(N_f-N_c)}x_i^{i-(N_f-N_c+1)}=\Lambda^{3N_c-N_f}
,~~\mbox{for $N_f-N_c<i\leq N_f$}.
\label{above}
\eeqa
Let us see how this curve behaves in the limit
$\Lambda^{3N_c-N_f}\to 0$ in each patch:

\noindent
($i\leq N_f-N_c$):~
Equation (\ref{below}) shows that
either $y_i$ or $x_i$ diverges. Namely, the curve $C$ goes away from this
patch.\\[0.1cm]
($i=N_f-N_c+1$):~
 Equation (\ref{above}) reads $y_i=\Lambda^{3N_c-N_f}$,
which becomes $y_i=0$ in the limit. Namely, the curve $C$ wraps once on the
$\CP^1$ cycle $C_i=C_{N_f-N_c+1}$.\\[0.1cm]
($i=N_f-N_c+2$):~
 Equation (\ref{above}) reads
$y_i^2x_i=\Lambda^{3N_c-N_f}$. As $\Lambda^{3N_c-N_f}$ is sent small,
for a fixed non-zero $y_i$, the $x_i$ value becomes small.
On the other hand,
for a fixed non-zero $x_i$, the two branches of $y_i$ value
both become small. In the limit $\Lambda^{3N_c-N_f}\to 0$,
the curve $C$ wraps once on $C_{i-1}=C_{N_f-N_c+1}$ as we have
seen, and twice on $C_i=C_{N_f-N_c+2}$.\\[0.1cm]
($i>N_f-N_c+1$ in general):~
 Equation (\ref{above})
shows that, in the limit,
the curve $C$ wraps $i-(N_f-N_c+1)$ times on $C_{i-1}$
and $i-(N_f-N_c)$ times on $C_i$.

Thus, in the strict $\Lambda^{3N_c-N_f}=0$ limit,
the component $C$ wraps on the $\CP^1$ cycles $C_1,\ldots,C_{N_f-1}$
with the following multiplicity\footnote{This argument was used in
\cite{HOO} to obtain this configuration for $N=1$ SQCD by
rotating the $N=2$ configuration by an angle $\mu$ and taking the
$\mu\to\infty$ limit.}
\beq
\underbrace{0,\ldots,0}_{N_f-N_c},1,2,\ldots,N_c-1.
\eeq

If we look at the last patch ($i=N_f$), we see that the curve wraps
$N_c$-times the locus of $y_{N_f}=0$.
This locus is the $x$ axis which looks like a semi-infinite
cigar. In particular, it appears that the curve becomes
infinitely elongated and the worldvolume theory
can no longer be considered as a four-dimensional theory.
However, we recall here that we are taking the limit $R\to 0$ at the
same time by keeping $\Lambda^{3N_c-N_f}\e^{L/R}=E^{3N_c-N_f}$ finite
for some $L$.
Then, the $v$ value starts growing exponentially
at some value of $x^6$ around $x^6=L$ in the scale set by $E$.
Beyond such values of $x^6$, gluons and gluinos have a large mass
compared to $E$, and does not contribute to the dynamics.

\begin{figure}[htb]
\begin{center}
\epsfxsize=6.0in\leavevmode\epsfbox{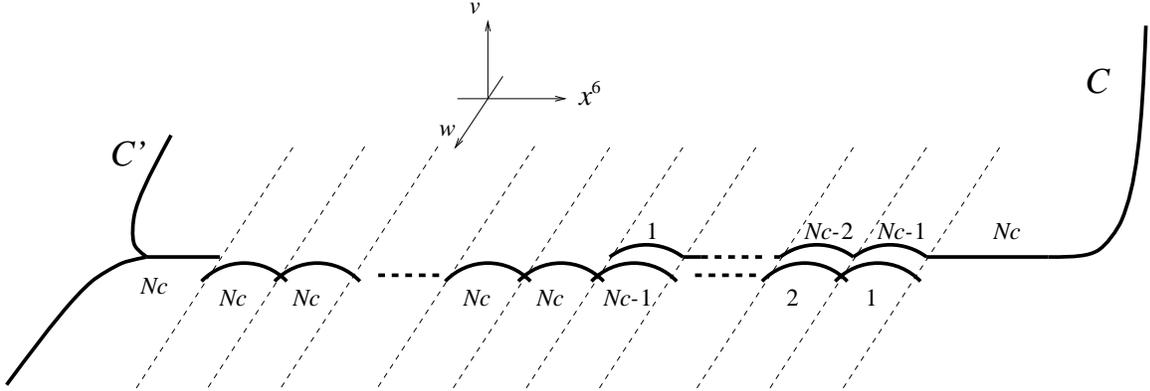}
\end{center}
\caption{The Electric Limit.
(The original $\CP^1$ components have
been moved from $\vec{w}=0$ in order to avoid confusion with
the new ones.)}
\label{elecfig}
\end{figure}

Together with the other infinite component $C^{\prime}$ and the
original $\CP^1$ components, we find that
the configuration of the fivebrane
looks as in Figure \ref{elecfig}.
In the Type IIA limit,
this is nothing but the configuration
of \cite{EGK} describing the $N=1$ SQCD with gauge group $SU(N_c)$
and $N_f$ flavors.

Actually, we have not carefully looked at the behaviour of the component
$C^{\prime}$ in the limit.
The component $C^{\prime}$
starts blowing up in the $w$ direction
at $x^6=-L^{\prime}$ where $L^{\prime}$ is given by
$c\, \e^{L^{\prime}/R}\sqrt{2L^{\prime}}^{N_f}
\sim E^{2N_c}$
(for simplicity we have put the D6-branes on top of each other).
If we fix the constant $c$ as in the discussion above,
$L^{\prime}$ approaches zero as $R\to 0$ and therefore
the component
$C^{\prime}$ becomes close to the D6-branes.
\footnote{It is interesting to note that
this limit leads to a configuration where an NS${}^{\prime}$
brane divides the D6-branes into two parts. This is the configuration
to which the chiral flavor symmetry is attributed in \cite{BH}.}
If we want to have a configuration as depicted in Figure \ref{elecfig},
we must take a limit $c\to 0$ such that
$c\, \e^{L^{\prime}/R}\sqrt{2L^{\prime}}^{N_f}\sim E^{2N_c}$
holds for fixed $L^{\prime}$ and $E$.
In such a limit, $y$ and $x$ degenerates
even away from the sixbranes and
the Taub-NUT space is no longer well-described by
the coordinates $(y_i,x_i)$ defined by (\ref{iyx})-(\ref{reliyx}).
However, we can rescale the coordinates so that the configuration
is still described by (\ref{below}) and (\ref{above}) with
$\Lambda^{3N_c-N_f}$ being replaced by $c^{-1}\Lambda^{3N_c-N_f}$.
Therefore,
as far as $c^{-1}\Lambda^{3N_c-N_f}$ becomes zero in the limit
(i.e. as long as $L^{\prime}$ does not exceed $L$),
the behaviour of the other component $C$ remains the same
and we obtain the configuration as depicted in Figure \ref{elecfig}.
Things are different if
$c\to 0$ is taken so that $c^{-1}\Lambda^{3N_c-N_f}$ does not become
zero; for example, if it blows up.
It will become clear what we will obtain in such a limit
once we consider the other limit, $\Lambda^{3N_c-N_f}\to\infty$
as we will do shortly.

The above configuration was obtained
in \cite{HOO} by rotating the configuration for
asymptotic free $N=2$ SQCD and therefore, strictly speaking,
one can claim only for $N_f<2N_c$ that it corresponds to SQCD.
However, it is clear from what has been observed (identification of
the gauge coupling constant and the existence of correct Type IIA limit),
that it applies also for $2N_c\leq N_f<3N_c$.
This is actually true even for larger number of flavors $N_f>3N_c$.
In either case, if $\Lambda^{3N_c-N_f}\ll E^{3N_c-N_f}$
the gauge theory on the brane
is weakly coupled at the energy $E<1/\elst\ll 1/R$
where extra degrees of freedom are heavy, and therefore the configuration
does indeed correspond to SQCD.
For $N_f>3N_c$, the limit $\Lambda^{3N_c-N_f}\ll
E^{3N_c-N_f}$ corresponds to $E\ll \Lambda$.
In fact, $\Lambda$ in such a case is considered as
the Landau pole of the infra-red free theory, and
hence at energies much smaller than $\Lambda$
we have weakly coupled gauge theory.

\subsection{The Magnetic Limit}

\newcommand{\tilw}{\widetilde{w}}
\newcommand{\tilv}{\widetilde{v}}

Let us consider the opposite limit $\Lambda^{3N_c-N_f}\to \infty$
and look at the configuration with respect to some fixed scales in the
$v$ and $w$ directions.

From (\ref{below}) and (\ref{above}),
we see that the component $C$ behaves
in the limit $\Lambda^{3N_c-N_f}\to\infty$ as follows:

\noindent
($i< N_f-N_c$):~
Equation (\ref{below}) shows that, in the limit, the curve $C$ wraps
$N_f-N_c+1-i$ times on $C_{i-1}$ and $N_f-N_c-i$ times on $C_i$.
\\[0.1cm]
($i=N_f-N_c$):~
Equation (\ref{below}), $x_i=\Lambda^{-(3N_c-N_f)}$,
becomes $x_i=0$ in the limit, showing that the curve $C$ wraps
once on $C_{i-1}=C_{N_f-N_c-1}$.\\[0.1cm]
($i>N_f-N_c$):~
Equation (\ref{above}) shows that either $y_i$ or $x_i$ diverges.
This means that the component $C$ goes away from this patch.

Thus, in the strict limit $\Lambda^{3N_c-N_f}=\infty$,
the component $C$ wraps on the $\CP^1$ cycles with the multiplicity
\beq
N_f-N_c-1,\ldots,2,1,\underbrace{0,\ldots,0}_{N_c}.
\eeq

If we look at the first patch ($i=1$), it appears that the component
$C$ wraps $N_f-N_c$ times on the $y$ axis (described by $x_1=0$)
and becomes infinitely elongated in the limit
$\Lambda^{3N_c-N_f}\to\infty$.
However, if we take also the limit $R\to 0$ at the same time by keeping
fixed $\Lambda^{-(3N_c-N_f)}\e^{L/R}$ for some $L$,
the value of $v$ starts growing exponentially at some value of
$x^6$ around $x^6=-L$.

\begin{figure}[htb]
\begin{center}
\epsfxsize=6.0in\leavevmode\epsfbox{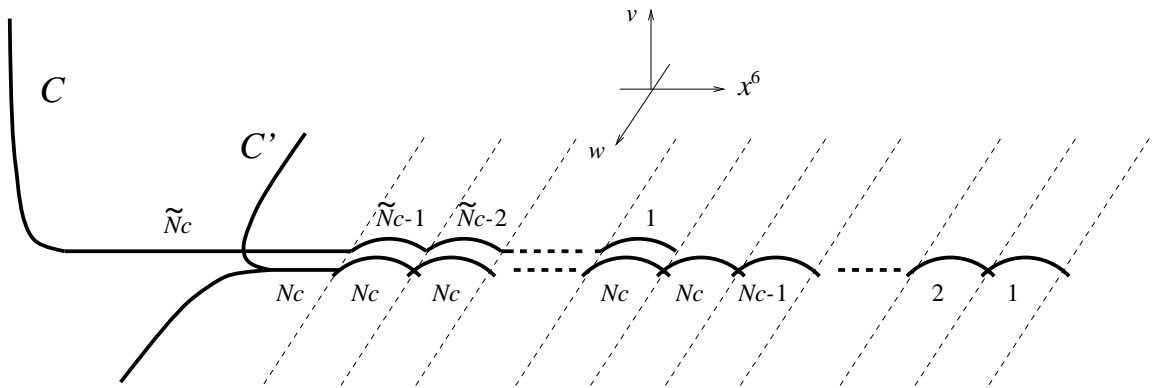}
\end{center}
\caption{The Magnetic Limit.
(The component $C^{\prime}$ has also been moved from $\vec{w}=0$
to avoid confusion with the component $C$.)}
\label{magfig}
\end{figure}

Together with the component $C^{\prime}$ and the original $\CP^1$ components,
we see that the fivebrane configuration in the
limit looks as in Figure \ref{magfig}.
Here again,
in order for the component $C^{\prime}$ to look as
in Figure \ref{magfig},
we need to tune the parameter $c$ as $c\to 0$ as indicated
in the electric limit. If we keep $c$ fixed, the component
$C^{\prime}$ approaches the sixbranes.

In the weakly coupled Type IIA limit,
the configuration in Figure \ref{magfig}
is nothing but the one of \cite{EGK} describing the magnetic theory.
In particular, we can recognize the $SU(\tilNc)$ gauge symmetry
in a part of the component $C$, namely, in the region sandwiched between
the part of $C$ with large values of $v$ (which is identified with
the NS 5-brane) and the part of $C^{\prime}$ with large values of $w$
(which is identified with the NS${}^{\prime}$ 5-brane).
The rest of the curve, to the right of the NS${}^{\prime}$
part of $C^{\prime}$,
are identified with the D4-branes
representing the singlet meson field.

Unlike in the brane motion of Type IIA configuration
\cite{EGK}, we do not have to turn on and off the
``Fayet-Iliopoulos parameter'' to avoid the
intersection of the NS and \NSp fivebranes.
The components $C$ and $C^{\prime}$
auotomatically avoid each other and are well-separated
in the process before taking the Type IIA limit.

\subsection*{\sl The Magnetic Coupling}

In order to measure the $SU(\tilNc)$ gauge coupling constant, we must be more
precise about the identification of the regions of $C$ and $C^{\prime}$
with the NS, D4 and NS${}^{\prime}$ branes.
For this, we need to set an energy scale $E$.

\vspace{-0.12cm}
We start with identifying
the $N_f$ D4-branes stretched between
\NSp and the left most D6-brane. We see from Figure \ref{magfig} that
$N_c$ of them come from $C^{\prime}$ and $\tilNc$ from $C$.
Motion of these D4-branes in the $w$-direction
corresponds to the fluctuation of the $N_f$ eigenvalues of the gauge
singlet meson.
Those from $C$ are stuck at $w=0$ but those from $C^{\prime}$ varies
as a function of $y$ as $w\sim y^{1/N_c}$.
The singlet meson is light compared to the mass scale $E$
when the separation of the eigenvalues are smaller than $E$.
Therefore, we may identify the $\tilNc$ of the D4-branes under question
as the part in $C^{\prime}$
with $|w|\leq \mu E$, where we have introduced a constant
mass scale $\mu$ to match the dimension.
In other words, it is the part with $|y|<|y^{\prime}(E)|$
where $y^{\prime}(E)=(\mu E)^{N_c}$.
For this region of $y$, the distance between the two components $C$ and
$C^{\prime}$ in the $v$ direction is
much smaller than their distance in the $w$ direction
in the weakly coupled Type IIA limit.
Therefore, we can identify the $N_f$ D4-branes as the region of the
components $C$ and $C^{\prime}$ with $|y|<|y^{\prime}(E)|$
(except the part of $C$ wrapping the $\CP^1$ cycles).

\vspace{-0.12cm}
We next distinguish the $\tilNc$
D4-branes from the NS 5-brane in the component $C$.
Since the scale in the $w$ direction is already set by $w\sim\mu E$,
the scale in the $v$-direction is set by $v\sim \mu^{-1}E^2$.
Thus, the boundary between the NS5-brane and the D4-branes should be
at $y=y(E):=\Lambda^{3N_c-N_f}(\mu^{-1}E^2)^{N_f-N_c}$.

Therefore, we can identify the $\tilNc$ D4-branes responsible for
the $SU(\tilNc)$ gauge dynamics as the region of the component $C$
with $|y^{\prime}(E)|<|y|<|y(E)|$.
In particular, the gauge coupling constant $\widetilde{g}(E)$
of the magnetic $SU(\tilNc)$ theory at the energy $E$ is given by
\beq
\e^{-1/\widetilde{g}^2(E)}
={y^{\prime}(E)\over y(E)}={\mu^{N_f}\Lambda^{-(3N_c-N_f)}\over
E^{3\tilNc-N_f}}\,.
\label{magcoup}
\eeq
The relation (\ref{magcoup}) implies that the dynamical scale
$\tiLambda$ of the magnetic theory is given by
\beq
\Lambda^{3N_c-N_f}\tiLambda^{3\tilNc-N_f}=\mu^{N_f},
\eeq
which is a familiar formula in field theory \cite{IS}.

\subsection*{\sl Holomorphic Description of the Magnetic Theory}

Under a suitable change of the coordinates,
we can express the curve purely in terms of the observables
of the magnetic theory.
From the discussion of the energy scales, it is natural to
change the coordinates $w$ and $v$ as
\beq
w=\mu\, \tilw,~~~v=\mu^{-1}\tilv.
\label{tilvw}
\eeq
Accordingly, we also change the coordinates $y$ and $x$ as
$y=\mu^{N_c}\widetilde{y}$,
$x=\mu^{-(N_c+N_f)}\widetilde{x}$.
Then, the complex structure of the
space-time is described by $\widetilde{y}\widetilde{x}=\tilv^{N_f}$,
and the curve is given by
\beq
C~\left\{
\begin{array}{l}
\widetilde{x}=\tiLambda^{3\tilNc-N_f}\tilv^{N_c}\\
\tilw=0,
\end{array}
\right.
~~~~~
C^{\prime}~
\left\{
\begin{array}{l}
\widetilde{y}=\prod_{i=1}^{N_c}
(\tilw-\widetilde{M}_i)\\
\tilv=0.
\end{array}
\right.
\eeq
In the above expression, $\widetilde{M}_i$ are the eigenvalues of the
singlet meson in the magnetic theory which are related to the electric
variables $M_i$ by
\beq
M_i=\mu\widetilde{M}_i\,.
\eeq
The fact that there are only $N_c$ parameters is consistent with
the fact that at most only $N_c$ eigenvalues of the singlet meson
can have non-zero vevs \cite{Seiberg}.
By the transformation properties
under the rotation in the 45 and 89 planes,
\footnote{We record the $U(1)_{45}$ and $U(1)_{89}$ charges
of the parameters and fields:
$$
\begin{array}{cccccccccccc}
&x(\widetilde{x})&y(\widetilde{y})&v(\tilv)&w(\tilw)&
\Lambda^{3N_c-N_f}&\tilQ Q&B\widetilde{B}&
\tiLambda^{3\tilNc-N_f}&q\widetilde{q}&b\widetilde{b}&\mu\\
U(1)_{45}&2N_f&0&2&0&2N_c-2N_f&0&0&2\tilNc&2&2\tilNc&0\\
U(1)_{89}&-2N_c&2N_c&0&2&2N_c&2&2N_c&-2N_c&0&0&0
\end{array}
$$}
it is natural to identify
$\tilv$ as the magnetic meson field $q\widetilde{q}$.
This leads to an interpretation of our claim that
the region of D4-branes in $C$ is in
$|v|<\mu^{-1}E^2$, namely in
$|\tilv|<E^2$; For $|\tilv|>E^2$, the gauge group is broken at an energy
above $E$ by the Higgs mechanism and does not contribute to the dynamics.
Also, the fact that there is no characteristic parameter
in the $v$-direction is consistent with the fact that the
vacuum expectation value of the magnetic meson is zero.
As in the electric theory,
by looking at the $U(1)$ charge, we may identify the magnetic baryon
vev with the
separation of the values of $\tiLambda^{3\tilNc-N_f}\widetilde{y}$
at $\tilv=\tilw=0$ between the two components $C$ and $C^{\prime}$:
\beq
b\,\widetilde{b}=\tiLambda^{3\tilNc-N_f}{\sl \Delta}\Bigl.
\widetilde{y}\,\,\Bigr|_{\tilv=\tilw=0}\,.
\eeq
In the above expression, $b\widetilde{b}$
is the combination $(1/\tilNc !)b_{i_1,\ldots,i_{\tilNc}}
\widetilde{b}^{i_1,\ldots,i_{\tilNc}}$
of the magenetic baryon fields which is invariant under
the action of the diagonal subgroup $U(N_f)$
of the chiral flavor symmetry group. 
By the equations defining $C$ and $C^{\prime}$, we have
\beq
b\,\widetilde{b}=(-1)^{N_c}\tiLambda^{3\tilNc-N_f}\prod_{i=1}^{N_c}
\widetilde{M}_i
\eeq
which agrees with the quantum modified constraint of the magnetic
theory. Actually, this can be considered as a consequence of the
relation $B\widetilde{B}:={\sl \Delta}y|_{v=w=0}
=(-1)^{N_c}\prod_{i=1}^{N_c}M_i$
corresponding to the classical constraint of the electric
theory under the map \cite{IS} of the observables
\beq
b\,\widetilde{b}=\mu^{-N_c}\tiLambda^{3\tilNc-N_f}B\widetilde{B},
\eeq
which follows from the coordinate
transformation $\widetilde{y}=\mu^{-N_c}y$.

\subsection*{\sl The Scale $\,\mu$}

We now determine the value of the scale $\mu$.
Let us go back to the problem of distinguishing
the D4-branes from the NS${}^{\prime}$ brane in the component $C^{\prime}$.
These $N_c$ D4-branes, together with the $\tilNc$ D4-branes from
the component $C$, are interpreted as representing
the $N_f$ eigenvalues of the
gauge singlet meson field, which is a chiral multiplet in the adjoint
representation of the diagonal flavor group $U(N_f)$.
Therefore, it is the same kind of problem as the one of section 2.1
distinguishing the D4-branes from the NS5-branes in the
configuration of the $N=2$ super-Yang-Mills theory.
As in that case, the D4-branes are identified as
the region where the configuration looks like
nearly coincident D4-branes compared to the length scale
$\elst^2E$.
Since the length in the $w$-direction is measured by
$|\elel^3{\sl\Delta}w|$, the $N_c$ D4-branes corresponds to
the region in $C^{\prime}$
with $|\elel^3w|<\elst^2E$, namely $|w|<{1\over R}E$.
This shows that
\beq
\mu={1\over R}.
\eeq

The space-time metric in the 45-89
directions, which is expressed at large $v\gg R/\elst^2$
as $|\elst^2\dd v|^2+|\elel^3\dd w|^2$,
is expressed as
$|\elel^3\dd \tilv|^2+|\elst^2\dd \tilw|^2$
in terms of the magnetic coordinates
$\tilv$ and $\tilw$ defined by (\ref{tilvw}).
Note the symmetry between the two expressions of the metric.

\newcommand{\tilZ}{\widetilde{Z}}
So far, we have been assuming $Z=1$ and also that the magnetic quarks
have canonically normalized kinetic term.
We can easily relax this conditon as follows.
We consider $Z$ to be general, and also the magnetic quarks to
have kinetic term $\tilZ (q^{\dag}q+\tilq\tilq^{\dag})$ at the energy $E$.
Then, the scales in the $\tilv$ and $\tilw$ directions at this
energy are set by
$\tilv\sim \tilZ^{-1}E^2$ and $\tilw\sim \tilZ E$.
Since the length scale in the $w$-direction is set by
$\elst^2 E$, we have
$\elst^2E=\elel^3Zw=\elel^3Z\mu\tilw$ and therefore
the scale $\mu$ is given by
$\mu=Z^{-1}\tilZ^{-1}{1\over R}$.
Then the space-time metric in the 45-89 direction at large $v$
is expressed as
$|\elst^2Z^{-1}\dd v|^2+|\elel^3 Z\dd w|^2
=|\elel^3\tilZ\dd \tilv|^2+
|\elst^2\tilZ^{-1}\dd \tilw|^2$.

\subsection{Deformations}

\subsubsection{Mass Perturbation of the Electric Theory}

~~~Let us consider turning on a mass term
of one of the quarks of the electric theory
\beq
W_{\it tree}=m\tilQ^{i_*}Q_{i_*}~~~\mbox{(no sum over $i_*$)}
\label{masseq}
\eeq
This corresponds in the brane picture to lifting one of the sixbranes
from $v=0$ to $v=-m$.
The W-boson stretched between the sixbrane and the D4-branes has a
mass $\elst^{-2}\Delta|x^4+ix^5|=Z^{-1}|m|$.
This is nothing but the quark mass under the identification of
$Z$ in (\ref{defvw}) as the normalization factor of the quark
kinetic term
$\int\dd^4\theta\,Z(Q^{\dag}Q+\tilQ\tilQ^{\dag})$.

Suppose we are lifting the $i_*$-th sixbrane (from
the left) with $1<i_*<N_f$.
Then, the two-cycles $C_{i_*-1}$ and $C_{i_*}$ defined as the
fibres over the segments $[\vec{x}_{i_*-1},\vec{x}_{i_*}]$
and $[\vec{x}_{i_*},\vec{x}_{i_*+1}]$ become tilted and are no longer
supersymmetric with respect to the supersymmetry preserved by
the fivebrane components wrapped on other cycles.
However, the cycle $\widetilde{C}$ defined as the fibres over
the straight segment $[\vec{x}_{i_*-1},\vec{x}_{i_*+1}]$
is in the line $v=0$ and is still supersymmetric.
This cycle is a smooth two-sphere with a neck in the middle
which has a circumference $\sim 4\pi\elst^2|m|$ if $\elst^2m\ll R$.
\begin{figure}[htb]
\begin{center}
\epsfxsize=6.0in\leavevmode\epsfbox{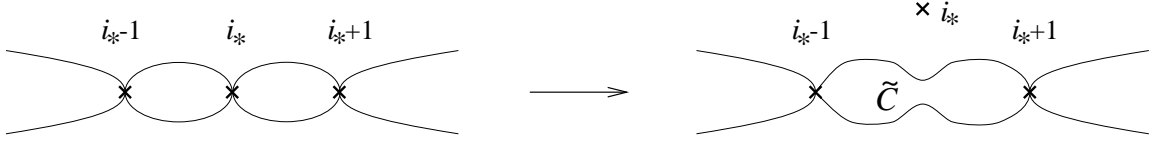}
\end{center}
\caption{Lifting the $i_*$-th Sixbrane}
\label{mass}
\end{figure}
What has been said applies also for $i_*=1,N_f$ with an
obvious modification.

The fivebrane configuration is supersymmetric after the deformation
only when the components wrapped on $C_{i_*-1}$ and
$C_{i_*}$ merge into components wrapped on
$\widetilde{C}$.
This is possible only if the numbers of components wrapped on
$C_{i_*-1}$ and $C_{i_*}$ are equal and their location in the
$x^{7,8,9}$ directions match.
Therefore we can lift the sixbrane only for $i_*=1,\ldots,N_f-N_c$.
In particular we cannot lift any one of the sixbranes when $N_f=N_c$.
This is actually consistent with field theory since giving a mass to
a quark would lead to SQCD with $N_c-1$ massless quarks
which has no supersymmetric stable vacuum \cite{ADS}.

The holomorphic description of the curve after lifting
the sixbrane remains the same as
(\ref{HOOconf}) where $x$ and $y$ now obey the relation
\beq
yx=v^{N_f-1}(v+m).
\eeq
The limit $m\to\infty$ exists if $\Lambda$ is sent to zero so that
$\Lambda_L^{3N_c-(N_f-1)}=m\Lambda^{3N_c-N_f}$ is kept fixed
and if $\hat{x}=m^{-1}x$ becomes a good coordinate.
In this case, the expression of the curve (\ref{HOOconf})
goes over to the one corresponding to the theory with $N_f-1$
flavors whose dynamical scale is $\Lambda_L$.

In terms of the real coordinates, $y$ and $x$ are expressed for
large $m$ as
\beqa
&&y=\hat{c}\times\e^{-(x^6/R+ix^{10})}\prod_{i\ne i_*}
\sqrt{|\vec{x}-\vec{x}_i|-(x^6-x^6_i)}\,.
\\
&&x=m\,\hat{c}^{-1}
\times\left({v\over |x^4+ix^5|}\right)^{N_f-1}
\e^{x^6/R+ix^{10}}\prod_{i\ne i_*}
\sqrt{|\vec{x}-\vec{x}_i|+(x^6-x^6_i)}\,,
\eeqa
where $\hat{c}=c\elst\sqrt{|m|/Z}$.
Therefore $\hat{x}$ (and also $y$) remains a good
coordinates in the limit $m\to\infty$
if $c$ is sent to zero so that $\hat{c}$ is kept fixed.

\begin{figure}[htb]
\begin{center}
\epsfxsize=2.0in\leavevmode\epsfbox{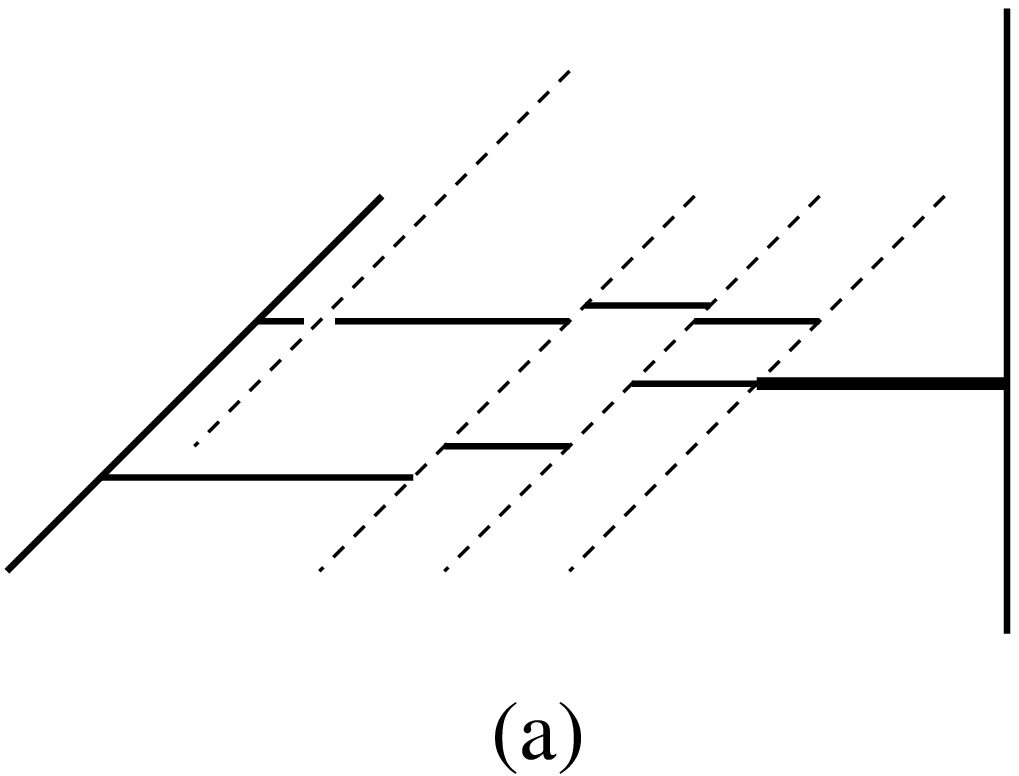}
\qquad\qquad\quad
\epsfxsize=2.25in\leavevmode\epsfbox{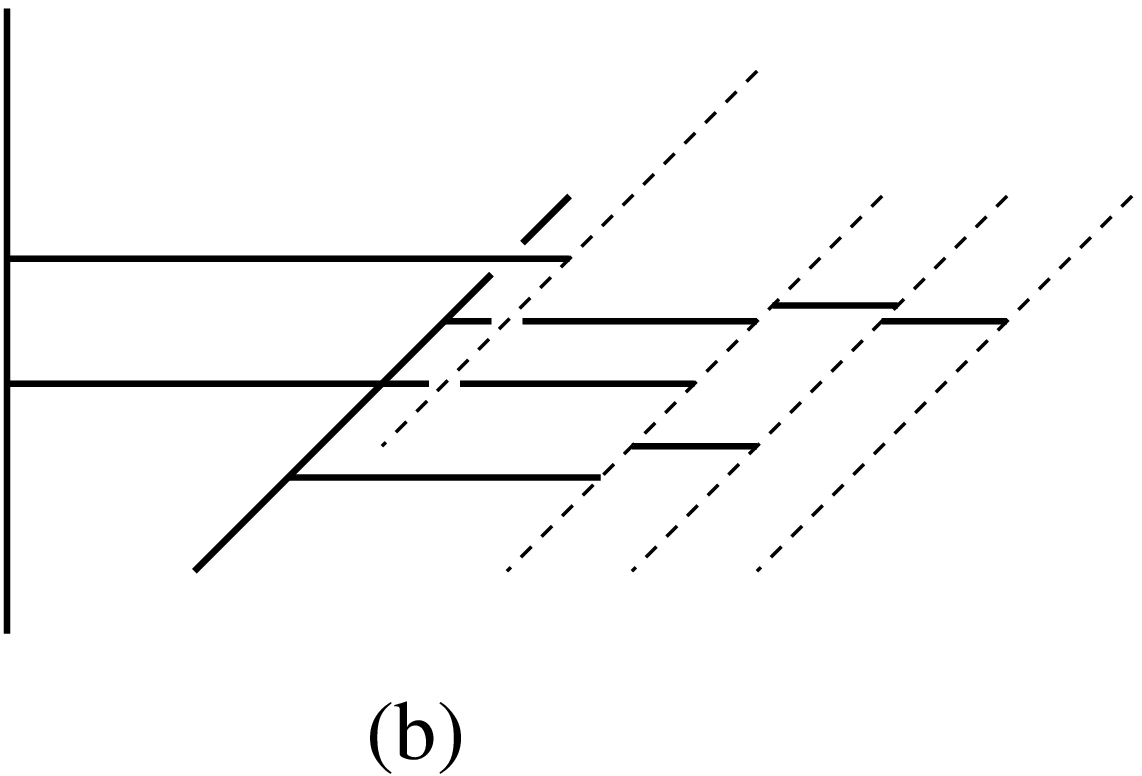}
\end{center}
\caption{The Deformation (\ref{masseq}) in the
case $N_c=2$, $N_f=4$.}
{\small The mass deformation in the electric theory (a)
corresponds to the Higgsing in the magnetic theory (b).}
\label{massEM}
\end{figure}

It is easy to see the consequence of lifting the sixbrane
in the Type IIA limits.
In the electric limit, it is just to lift a D6-brane with certain
rejoining of the D4-branes (see Figure \ref{massEM}).
In the magnetic limit,
a D4-brane ending on the lifted sixbrane is created
(see Figure \ref{massEM}).
It is at $\widetilde{v}=-\mu m$ in the magnetic coordinate.
This corresponds to the Higgsing of the magnetic theory,
$q^{i_*}\tilq_{i_*}=-\mu m$, induced
by the superpotential $W_{\it mag}={1\over \mu}Mq\tilq
+mM^{i_*}_{i_*}$.
This is the brane realization of the correspondence of
mass-deformation of the electric theory (making the coupling stronger)
and the Higgsing of the magnetic theory (making the coupling weaker).

\subsubsection{Higgsing the Electric Theory}

~~~It is possible to have a supersymmetric configuration if we lift
{\it all} of the sixbranes from $v=0$:
\beq
yx=\prod_{i=1}^{N_f}(v+m_i),~~~m_i\ne 0.
\eeq
This corresponds in field theory to giving bare mass to all of the
quarks, and the low energy physics is that of $SU(N_c)$
super-Yang-Mills theory with the dynamical scale $\Lambda_L$
given by $\Lambda_L^{3N_c}=m_1\cdots m_{N_f}\Lambda^{3N_c-N_f}$
which exhibits chiral symmetry breaking $\Z_{2N_c}\to\Z_2$.
The baryon operators have zero vevs
but the meson vev is non-zero,
$M={\rm diag}(\Lambda_L^3/m_i)$.
The brane configuration as obtained in \cite{HOO,BIKSY}
consists of a single component
\beqa
vw&=&\Lambda_L^3,\\
x&=&\Lambda^{-3N_c+N_f}v^{N_c}.
\eeqa
Indeed, there are $N_c$ possible configurations
corresponding to the $N_c$ solutions for $\Lambda_L^3$.
The result applies also to the case $N_f<N_c$.
The relation between $y$ and $w$ is
$y=w^{N_c-N_f}\prod_{i=1}^{N_f}(w+\Lambda_L^3/m_i)$
and the eigenvalues of the meson matrix appears as
the root of $y=0$.
In the case $N_f<N_c$, the meson vev $M=\Lambda_L^3/m$
follows from the dynamical generation of the superpotential
\cite{ADS} in the field theory side.
Thus, one can say that
the brane configuration encodes an information about
the superpotential generation.
There is also a similar interpretation for higher flavor cases
$N_f>N_c$.

Let us consider the limit where $N_c$ of $m_i$ approaches zero
as $m_i=\alpha \tilde{m_i}$, $\alpha\to 0$ ($i=N_f-N_c+1,\ldots,N_f$)
with $\tilde{m}_i$ kept finite.
Then, $\Lambda_L^3/m_i$ approaches zero for
$i=1,\ldots,N_f-N_c$, but it remains the same for
$i=N_f-N_c+1,\ldots, N_f$.
In field theory, this leads to a vacuum of SQCD with $N_c$ massless
and $N_f-N_c$ massive quarks given by
\beq
M={\rm diag}(0,\ldots,0,M_1,\ldots,M_{N_c}),~~~
B=\widetilde{B}=0,
\label{landing}
\eeq
where $M_i=\Lambda_L^3/m_{N_f-N_c+i}$.
At this stage, we can send $m_1,\ldots, m_{N_f-N_c}$ to zero
keeping fixed $M_i$, which yields a vacuum of
SQCD with $N_f$ massless quarks with the meson and baryon vevs as in
(\ref{landing}).
\begin{figure}[htb]
\begin{center}
\epsfxsize=4.5in\leavevmode\epsfbox{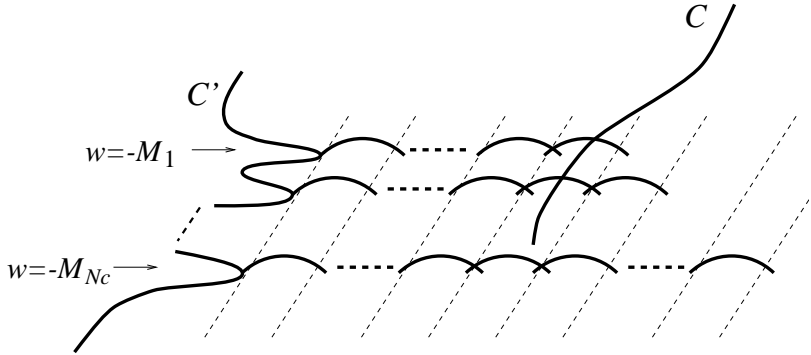}
\end{center}
\caption{The Configuration Corresponding to the Vacuum (\ref{landing})}
\label{land}
\end{figure}
In the brane picture, the corresponding procedure leads to the
configuration as depicted in Figure \ref{land},
where the $\CP^1$ components make rows at
$w=-M_1,\ldots,-M_{N_c}$.

\subsection*{\sl Decoupling}

Let us consider sending one of the meson eigenvalues,
say $M_{N_c}$ in (\ref{landing}), to infinity:
\beq
M_{N_c}\longrightarrow \infty.
\label{Decoup}
\eeq
In field theory,
this leads to $SU(N_c-1)$ SQCD with $N_f-1$ flavors
when $\Lambda$ is tuned so that
$M_{N_c}^{-1}\Lambda^{3N_c-N_f}$ is kept fixed.
The dynamical scale $\hat{\Lambda}$
of this theory is given by
$\hat{\Lambda}^{3(N_c-1)-(N_f-1)}
=M_{N_c}^{-1}\Lambda^{3N_c-N_f}$.

In the brane picture,
this corresponds to sending the $N_c$-th row at $w=-M_{N_c}$
to infinity.
One can also send the right-most D6-brane to the infinity
$x^6_{N_f}\to +\infty$ at the same time.
Then, it is appropriate to use the coordinates
$\hat{y}=M_{N_c}^{-1}y$ and $\hat{x}=M_{N_c}v^{-1}x$
which are expressed in the limit as (\ref{defy}) and (\ref{defx})
where $N_f$ and $c$ are replaced by $N_f-1$ and
$\hat{c}=M_{N_c}^{-1}\sqrt{2x^6_{N_f}}c$.
In terms of these coordinates, the expression of the curve
(\ref{HOOconf}) goes over to the one with
$N_c,N_f,\Lambda$ being replaced by
$N_c-1,N_f-1,\hat{\Lambda}$.
The number of components wrapping $C_i$ has reduced by one
for $i=1,\ldots,N_f-2$ while the last $\CP^1$ cycle $C_{N_f-1}$
has become infinitely elongated and no fivebrane wrapps on it.

\begin{figure}[htb]
\begin{center}
\epsfxsize=6in\leavevmode\epsfbox{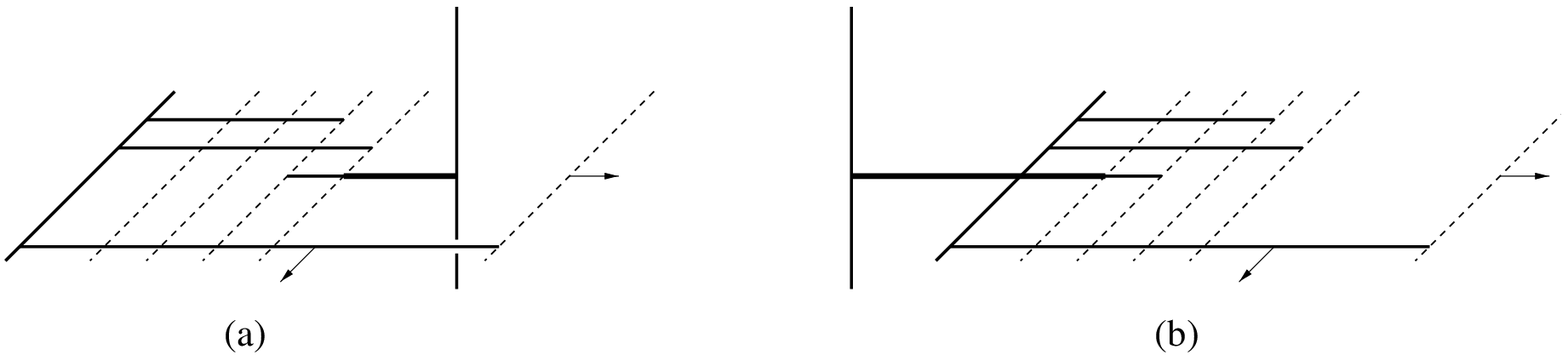}
\end{center}
\caption{The Decoupling (\ref{Decoup}).}
{\small Higgsing in the electric theory (a)
corresponds to giving a mass to a quark in the magnetic theory (b).}
\label{Higgsel}
\end{figure}
In Figure \ref{Higgsel}, we depict the two Type IIA limits of the above
process.
In the magnetic limit, the color remains the same but
one quark has decoupled, $N_f\to N_f-1$,
because of the large mass given by the singlet meson vev.

\subsection{Brane Move and Universality Class of the Worldvolume
Theory}

In this section, we have seen that the two
Type IIA brane configurations
describing the electric and magnetic theories
of \cite{Seiberg} arize from a common configuration of the fivebrane
in $M$ theory.
The space-time and the configuration depend on, among others,
the parameters $R$ and $\Lambda$ or $\tiLambda$ which are related
by $\Lambda^{3N_c-N_f}\tiLambda^{3\tilNc-N_f}\sim R^{-N_f}$.
The electric limit arizes by first sending
$\Lambda^{3N_c-N_f}\ll E^{3N_c-N_f}$
in the eleven-dimensional supergravity region $R\gg\elel$,
where $E$ is an energy scale determining
the unit of lengths in the $x^{4,5,7,8,9}$
directions,
and then scaling down to the weakly coupled
Type IIA string theory region $R\ll\elel$,
$E<M_{\it st}$ with the ratio $\Lambda/E$ kept fixed.
The magnetic limit arizes by first sending
$\tiLambda^{3\tilNc-N_f}\ll E^{3\tilNc-N_f}$
in the eleven-dimensional supergravity region
and then going down to the weakly coupled Type IIA region
with the ratio $\tiLambda/E$ kept fixed.
The two \MT configurations before the Type IIA limit are
smoothly interpolated without a change in the complex structure.

There have been a lot of works concerning the interpolation of the
electric and magnetic configurations by brane moves
within the weakly coupled Type IIA region \cite{GK}.
A move of branes, when it corresponds to a change
of parameters of the gauge theory,
changes the physics of the worldvolume theory.
If the move corresponds to a change of parameters
that have no counterpart in four-dimensional gauge theory,
then, it is believed that it does not change the physics.
An example of such ``irrelevant moves'' is 
changing the position of D6-branes in the $x^6$-direction
relative to each other or with respect to the NS5 and D4-brane system
(this fails to be irrelevant in some cases,
as we will mention shortly).
Another example is changing the relative separation of the
two NS5-branes, which corresponds to a change of the gauge coupling
constant. 
When the gauge theory is asymptotic free, or more generally, when
the gauge coupling runs at high energy, this just corresponds to
a change of the overall scale of the theory and is irrelevant
in the infra-red.

Indeed, what is done in the literature \cite{GK}
is to utilize the latter move to derive the $N=1$ duality.
However, in the course of interpolating the electric and magnetic
configurations by this move, there is a point at which
the NS5-brane intersect with the other NS5-brane or the D6-branes.
Thus, this brane move looks as a singular process and
it is not obvious whether we can expect that
the universality class of the
theory remains the same.
Actually it was proposed in the very first literature \cite{EGK}
that the singularity in the process may be avoided
by another move which corresponds to turning on
the Fayet-Iliopoulos
parameter of a $U(1)$ factor of the gauge group.
It is doubtful whether the gauge group has an $U(1)$
factor or not. However, even if there is a $U(1)$ factor,
turning on the FI term moves the theory away from the origin of
the moduli space at which most interesting physics is concentrated.
In addition, when the gauge group is symplectic
or orthogonal groups, which can be realized by introducing
an orientifold-plane, there is no room to turn on the FI parameter
and the singularity seems unavoidable \cite{EJS,EGKRS}.

What we have done in this section is related to this Type IIA brane
move since we have changed the parameter $\Lambda$ corresponding to
the dynamical scale of the gauge theory.
It can be considered as a resolution of the singularity
of the Type IIA brane move by going to $M$ theory.
The brane move in the supergravity region is completely smooth
and we do not have to turn on the FI parameter (which is actually
impossible since we are considering the the gauge symmetry 
to be $SU(N_c)$).
We will see in the following sections that the method
also applies effectively to the case of symplectic or
orthogonal gauge groups where FI term never exists:
the singularity which is inevitable in the Type IIA
move is resolved by going to $M$ theory.
Also, we have mainly considered the case where
the quark mass terms are turned off. The mass term, as the FI term,
would take the theory away from the interesting point.

Does the universality class of the theory
remain the same?
The most subtle point is of course the strong string coupling limit
$\gst\gg 1$ at the starting point of the interpolation, and
the limit $R\ll\elel$ in the final stage.
As $\gst$ is increased, degrees of freedom other than the gauge
fields starts strongly coupled
and the theory is no more a simple gauge theory.
Nevertheless, it is true that
a four-dimensional supersymmetric field theory
exists in the infra-red limit at an arbitrary values of
$\gst$: No matter how large it is, if one look at the configuration
from an extremely long distance, it looks the same as the starting
Type IIA configuration as far as we scale up everything as
indicated above.
The universality class of a theory is a discrete notion unless
there is a parameter coupled to a marginal operator.
In supersymmetric theories,
such a parameter is usually a complex parameter
that can enter into the superpotential.
In the present case, the parameters we are changing
are $R$ and $c$ which are independent real
parameters.\footnote{The situation would be different if one other
direction of the space-time, say $x^7$, is compactified on a circle.
Then, its radius can combine with the parameter $R$ to make a
complex parameter and the theory can continuously depend on it.}
Therefore, these cannot
enter into the superpotential but only in the K\"ahler potential.
Then, as long as the process of interpolation
is smooth, it is highly possible that there is no singular change
in the K\"ahler potential, and we can expect that
the universality class remains the same.
The fact that we can
identify the gauge coupling constant that fits with
NSVZ exact beta function suggests
that this worldvolume theory is smoothly connected to the
ordinary SQCD or its magnetic dual.
Therefore, there is a high possibility that the universality class
is constant throughout the process of interpolation.

\subsection*{\sl A General Condition}

One thing we are really interested in is under what kind of
brane move the universality class of the worldvolume theory
remain unchanged.
From what we have argued, we propose the following general
condition:
Two theories that are realized by Type IIA brane
configurations are equivalent in the infra-red limit
when they are smoothly interpolated by a family of
configurations of $M$ theory fivebrane whose complex structure
is constant through the process of interpolation,
where the family is parametrized by {\it real parameters}.

\begin{figure}[htb]
\begin{center}
\epsfxsize=3.5in\leavevmode\epsfbox{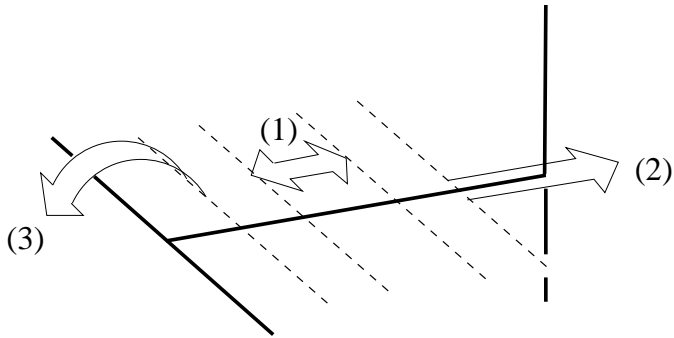}
\end{center}
\caption{Moving D6-branes}
\label{ILLU}
\end{figure}

For illustration, let us consider moves of D6-branes
of the electric Type IIA configuration in the
$x^6$-direction (see Figure \ref{ILLU}).

\noindent
(1)
Suppose we change the separation of two neighboring D6-branes,
which looks like a smooth move even in Type IIA region.
This corresponds to simply changing the size of the
$\CP^1$-cycle in the \MT configuration, and hence is a smooth
deformation of the curve with no change of the complex structure.
The size of $\CP^1$ is of course a real parameter. (This again fails
when we compactify another direction on a circle; the period of
the three-form in \MT becomes a complex partner.)
Therefore, it does not change the universality class.
Indeed, the K\"ahler metric of the Higgs branch
of $N=2$ theories was computed in \cite{DHOO}.
The metric explicitly depends on the separation of sixbranes,
but the relevant information --- the
information about the singularity --- is independent.
(In this particular example, it can explicitly be shown
that the dependence on the separation
disappears in the weakly coupled Type IIA limit $R/\elel\to 0$.)

\noindent
(2)
Let us next consider moving the right-most D6-brane to the right
of the NS5-brane. 
Then, the D6-brane necessarily intersect
with the NS5-brane and a single D4-brane stretched between them
is created. This looks like a singular process.
In \MT configuration, this corresponds to changing the real parameter
$c$ in the definition (\ref{defy}) of the complex
coordinates $y$ and $x$, and hence is a smooth deformation
of the curve without any change in the complex structure.
The brane ``creation'' in Type IIA simply corresponds to
an approximate wrapping of the fivebrane on the $\CP^1$
cycle, and is indeed a smooth process. Nothing is actually
created, and there is no singularity nor topology change.
Therefore, it does not change the universality class.

\noindent
(3)
Finally, let us move the left-most D6-brane to the left of the
NS${}^{\prime}$5-brane.
If we keep the D6-brane at $x^{4,5}=0$ through the process,
it must coincide with the NS${}^{\prime}$5-brane
and looks singular. Indeed, an
analogous move in \MT is singular as well.
However, one could move the D6-brane
around the NS${}^{\prime}$5-brane
by turning on and off the mass parameter. Namely,
first lift the D6-brane to $x^{4,5}\ne 0$,
move it to the left in the $x^6$-direction,
and then move it down to $x^{4,5}=0$.
This looks equally smooth compared to the move used in
``deriving'' the $N=1$ duality within the Type IIA
region: turning on and off the
FI parameter.
It is possible to do the similar move in the
\MT configuration. 
However, it is actually not a smooth move and it also changes the
complex structure. The singularity occurs at the very beginning of the
move. When the mass is turned on the topology of the configuration
changes, as we can see by looking at Figure \ref{mass}.
After that, the configuration is smoothly deformed without
any change of the complex structure.
Therefore, the total move may change the universality class.
Actually, the configuration after the move is smoothly interpolated
to the configuration with one less flavors
without change of the complex structure,
by sending the D6-brane to $x^6\to-\infty$.
Thus, the universality class after the move
is the same as the one of the theory with $N_f-1$ flavors.

\section{Symplectic Groups}

In this section, we generalize the above consideration to
SQCD with symplectic gauge groups. We show that
the fivebrane configuration has two Type IIA limits
corresponding to the electric and magnetic descriptions.
Also, we show that
the fivebrane configuration proposed in \cite{DSB}
for the dynamical supersymmetry breaking
model of \cite{IY,IT} (IYIT model)
has the correct Type IIA limit.

\subsection{SQCD with Symplectic Gauge Group}

In order to realize symplectic gauge group,
we consider \MT background that reduces in the Type IIA limit
to orientifold four-plane.
In particular,
we consider \MT on $\R^4\times M^7/\Z_2$
where $M^7$ is a product of the Taub-NUT space
of $A_{2N_f-1}$ type (described by $y,x,v$ as in the previous
sections) and $\R^3$ (parametrized by $x^7$ and
$w\propto x^8+ix^9$),
and the $\Z_2$ acts as $y\to y$, $x\to x$, $v\to -v$ and
$x^{7,8,9}\to -x^{7,8,9}$.
The Taub-NUT space is covered with $2N_f$ coordinate patches
as explained in the previous section,
and the $\Z_2$ acts on the coordinates as $(y_i,x_i)\to
((-1)^{i-1}y_i,(-1)^{i}x_i)$. 
In particular, it fixes the $\CP^1$ cycles $C_{2i}$
point-wisely, but acts on $C_{2i+1}$ as the $\pi$-rotation
around the intersection points with $C_{2i}$ and $C_{2i+2}$.

\subsection*{\sl The Fivebrane Configuration}

The fivebrane configuration
for $Sp(N_c)$ SQCD with $N_f>N_c$ flavors can be
obatined by rotating
the $N=2$ configuration of \cite{H2}.
\footnote{This was essentially done in \cite{DSB}, but
there was a subtle error in the starting $N=2$ configuration.
Here we start with the correct one \cite{H2}
and present the correct result for $N=1$ SQCD.}

\begin{figure}[htb]
\begin{center}
\epsfxsize=5.8in\leavevmode\epsfbox{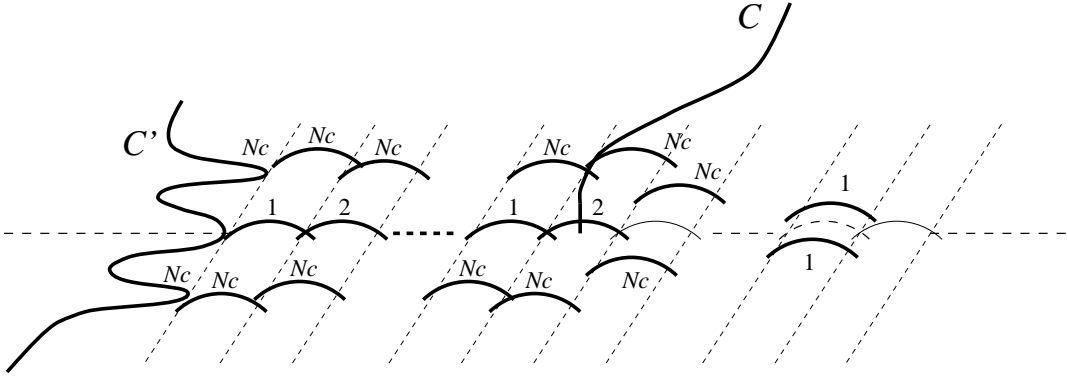}
\end{center}
\caption{The $M$ Theory Configuration of $Sp(N_c)$ SQCD.}
{\small Half of the component $C$ is not written here in order
to avoid complication. There are $2N_f$ sixbranes.
The dashed horizontal lines and curves correspond to
the $\Z_2$ fixed point set whereas the thin curves stand for $\CP^1$
cycles on which the $\Z_2$ acts as the $\pi$-rotation.}
\label{Spconfig}
\end{figure}

The fivebrane consists of two infinite components in $x^7=0$
\beq
C^{\prime}~
\left\{
\begin{array}{l}
y=\prod_{i=1}^{N_c+1}(w^2-M_i^2)\\
v=0,
\end{array}
\right.
~~~~~
C~\left\{
\begin{array}{l}
x=\Lambda^{-(6(N_c+1)-2N_f)}v^{2(N_c+1)}\\
w=0,
\end{array}
\right.
\label{Spconf}
\eeq
and several $\CP^1$ components wrapping the cycles
$C_1,\ldots,C_{2N_f-1}$.
The number of fivebranes wrapping the cycle $C_i$
is $2N_c+1$ for odd $i$ and
$2N_c+2$ for even $i$ if $1\leq i\leq 2N_f-2N_c-2$,
and it is $2[(2N_f-i)/2]$ for larger $i$,
$2N_f-2N_c-1\leq i\leq 2N_f-1$.
The infinite component $C$ intersects
transversely with the $y$-axis
at $y=\Lambda^{4(N_c+1)}$ if $N_f=N_c+1$ while it intersects with
the $\CP^1$ cycle $C_{2N_f-2N_c-2}$ if $N_f>N_c+1$.
In \cite{H2}, it was shown that
the Dirac quatization condition is not satisfied
when the fivebrane in $\R^5/\Z_2$ orbifold
intersects with the $\Z_2$ fixed plane
transversely in a four-dimensional factor of the space-time.
In other words, such an intersection, which we call
a {\it t}-configuration \cite{DSB}, is forbidden.
In order to avoid a {\it t}-configuration,
for $N_f=N_c+1$,
the intersection points of $C$ and $C^{\prime}$
with the $y$-axis in $x^{7,8,9}=0$ must coincide, which requires 
$\prod_{i=1}^{N_c+1}(-M_i^2)=\Lambda^{4(N_c+1)}$.
For $N_f>N_c+1$, since the number of fivebranes wrapping
$C_1$ is odd, the component $C^{\prime}$ must intersects with
the $y$-axis in $x^{7,8,9}=0$ at $y=0$,
which means that at least one $M_i$ must be
zero. Also, in the latter case, 
at least one pair of the $\CP^1$ components wrapping $C_{2i}$
must be at $x^{7,8,9}=0$ for each $1\leq 2i\leq 2N_f-2N_c-2$
(see Figure \ref{Spconfig}).

The parameters $\pm M_i$ in (\ref{Spconf})
are interpreted as the eigenvalues of the
vev of the meson matrix $M^{ij}=Q^i_aJ^{ab}Q_b^j$
($J^{ab}$ is the $Sp(N_c)$ invariant skew-symmetric form)
and the above constraint on $M_i$ is equivalent to the quantum
constraint on $M^{ij}$ in field theory.
The number of $\Z_2$ invariant
deformations of the curve subject to
the constraints to avoid {\it t}-configurations
agrees with the dimension $4N_fN_c-(2N_c^2+N_c)$
of the moduli space of vacua of the field theory.

The complex coordinates $y,x,v$ and $w$ are related to the real
coordinates $x^{\mu}$ as (\ref{defy}), (\ref{defx}) and
(\ref{defvw}) where $N_f$ is now replaced by $2N_f$.
In addition to the parameter and observables of
SQCD, the configuration depends on two real parameters 
$cZ^{2N_c+2}\e^{-\sum_ix^6_i/R}$
and $R$ as well as the separation between the
sixbranes.

\subsection*{\sl The Electric Limit}

\newcommand{\tilO}{\widetilde{\rm O4}^+}

Let us consider the limit 
$\Lambda^{3(N_c+1)-N_f}\ll E^{3(N_c+1)-N_f}$
for a fixed energy scale $E$ and
look at the fivebrane
with respect to the scale set by $v\sim E$, $w\sim E^2$.
The component $C$ is sent to the right and wraps on the $\CP^1$ cycles
$C_{2N_f-2N_c-1},\ldots, C_{2N_f-1}$ with multiplicities
$1,2,\ldots, 2N_c+1$.
It also wraps on the $x$-axis with mulitplicity $2N_c+2$
and then blows up in the $v$-direction at
$y(E)=\Lambda^{6(N_c+1)-2N_f}E^{2N_f-2N_c-2}$.
The component $C^{\prime}$ wraps on the $y$-axis and blows up in the
$w$-direction at $y^{\prime}(E)=E^{4(N_c+1)}$.
The two regions with large $v\gg E$ and large $w\gg E^2$
are interpreted as the NS5 and NS${}^{\prime}$5-branes
in the Type IIA limit.
Between them, the fivebrane wraps
$2N_c+2$ or $2N_c+1$ times on the eleventh direction
--- $2N_c+2$ on cycles point-wisely $\Z_2$ invariant
and $2N_c+1$ on cycles on which $\Z_2$ acts as the $\pi$-rotation.
In the weakly coupled Type IIA limit $R\ll \elel$,
a pair of fivebranes wrapping each of the $\Z_2$ fixed cycles
($C_{2i}$ and the $y$- and $x$-axis
at $x^{7,8,9}=0$) is identified with
the orientifold four-plane of $Sp$-type
with trivial RR $U(1)$ gauge field (which we denote by
O4${}^+$).
Similarly, a single fivebrane wrapped on each
of the cycles $C_{2i+1}$ at $x^{7,8,9}=0$
on which $\Z_2$ acts as the
$\pi$-rotation is identified with the
$Sp$-type O4-plane with a non-trivial RR Wilson line \cite{H2}
(which we denote by $\tilO$).
The rest of the fivebrane, wrapping $2N_c$-times on the
eleventh direction, are interpreted as $2N_c$ D4-branes.
These together yield $Sp(N_c)$ gauge symmetry.

The gauge coupling constant $g(E)$ at the energy $E$
can be read as in the $SU(N_c)$ case. It is given by
\beq
\e^{-2/g^2(E)}={y(E)\over y^{\prime}(E)}
=\left({\Lambda\over E}\right)^{6(N_c+1)-2N_f}.
\eeq
The factor of two can be explained as in \cite{LLL} as the consequence of
a particular embedding of $Sp(N_c)$ to $SU(2N_c)$.

As a summary, we depict in Figure \ref{Spelec} (e)
the electric Type IIA limit.
Note that the dimension of the moduli space can be counted also
by applying the ``s-rule'' found in \cite{H2} to
this configuration.

\subsection*{\sl The Magnetic Limit}

Let us next consider the opposite limit
$\Lambda^{3(N_c+1)-N_f}\to \infty$.
The component $C$ is sent to the left and wraps on the $\CP^1$ cycles
$C_{1},\ldots, C_{2N_f-2N_c-3}$ with multiplicities
$2N_f-2N_c-3,\ldots,2,1$.
It also wraps $2N_f-2N_c-2$ times
on the $y$-axis and starts blowing up in the $v$-direction
at some value of $y$.
This value of $y$ is much larger than
the value where the component $C^{\prime}$ starts blowing up in the
$w$-direction with respect to some scale which wll be specified
momentarily.
Between the two regions with large $v$ and $w$,
the fivebrane wraps $2N_f-2N_c-2$ times
on the eleventh direction.
In the Type IIA limit,
two of them are interpreted as providing a
charge $+2$ to the orientifold fixed plane
(consituting the O4${}^+$-plane)
while the rest, $2\tilNc=2N_f-2N_c-4$ of them,
are considered as $2\tilNc$ degenerate D4-branes.
These together yield $Sp(\tilNc)$ gauge symmetry.
We can measure the gauge coupling at an energy $E$
by setting
the corresponding length scales in the $v$ and $w$ directions.
As in $SU(N_c)$ case, these length scales are set by
$w\sim \mu E$ and $v\sim\mu^{-1}E^2$ where $\mu$ is
a constant mass scale,
and the $Sp(\tilNc)$ gauge coupling $\widetilde{g}(E)$
is measured as
\beq
\e^{-2/\widetilde{g}^2(E)}=
{\mu^{2N_f}\Lambda^{-6(\tilNc+1)+2N_f}\over E^{6(\tilNc+1)-2N_f}}.
\eeq
This yields the standard relation
$\Lambda^{6(\tilNc+1)-2N_f}\tiLambda^{6(\tilNc+1)-2N_f}
=\mu^{2N_f}$
between the dynamical scales of
the electric and magnetic theories \cite{IP}
(up to the sign ambiguity).

On the right of the region of $C^{\prime}$ with large $w\gg\mu E$
(which is identified as the NS${}^{\prime}$5-brane),
we have components of the fivebrane wrapping
several times on the eleventh direction.
The fivebrane wraps $2N_f$-times in the region sandwiched between
the NS${}^{\prime}$5-brane and the left-most D6-brane,
while it wraps $2[i/2]$-times on the $\CP^1$ cycle $C_{2N_f-i}$.
In the weakly coupled Type IIA limit, these are all interpreted as
the D4-branes stretched between the NS${}^{\prime}$ and the left-most
D6-brane or between neighboring D6-branes;
the number of times the fivebrane wraps is equal to
to the number of stretched D4-branes.
The $\Z_2$ fixed cycle $C_{2i}$ at $x^{7,8,9}=0$ is
identified with the orientifold four-plane of $SO$-type
(O4${}^-$-plane) whereas the $\Z_2$ invariant but non-fixed
cycle $C_{2i+1}$ at $x^{7,8,9}=0$
is identified with the $SO$-type O4-plane with a single D4-brane
stuck on it (O4${}^0$-plane) which has a non-trivial RR Wislon
line \cite{H2}.
The stretched D4-branes represent the degrees of freedom
corresponding to
the gauge singlet meson field of the magnetic theory.
Indeed, the number of possible deformations of these D4-branes
is
\beq
N_f+2\sum_{i=1}^{2N_f-1}[i/2]
={2N_f(2N_f-1)\over 2},
\eeq
which is equal to the number of independent components
of the gauge-singlet meson field
$\tilde{M}^{ij}=-\tilde{M}^{ji}$.

\begin{figure}[htb]
\begin{center}
\epsfxsize=6.2in\leavevmode\epsfbox{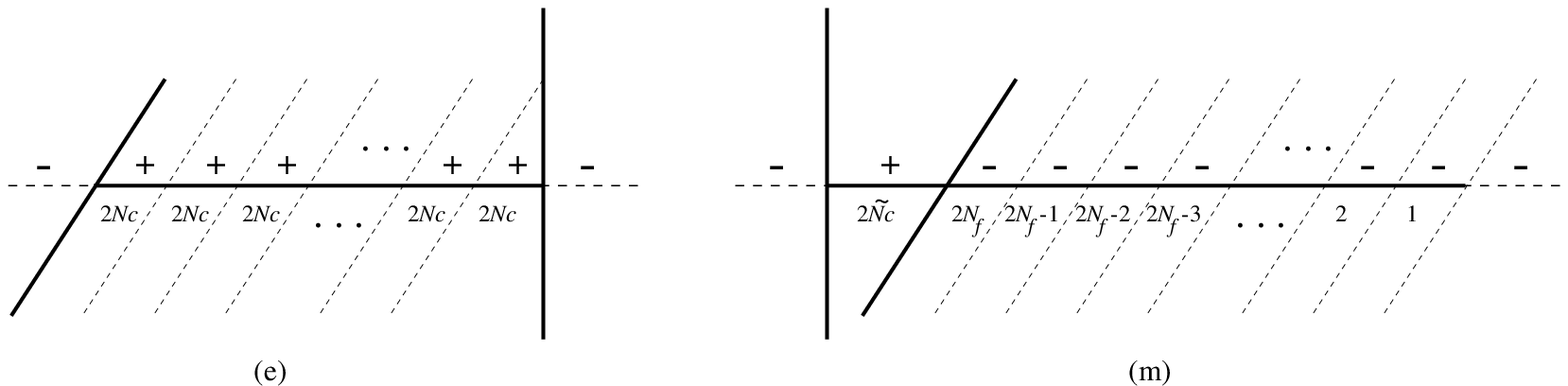}
\end{center}
\caption{The Electric (e) and Magnetic (m) Limits 
(Symplectic Groups).}
{\small 
``$+$'' and ``$-$'' stand for $Sp$-type and $SO$-type
O4-planes respectively. The numbers below indicate
the number of stretched D4-branes. We omit indicating the
RR Wilson line (it is described in the main text). 
There are $2N_f$ D6-branes in each.}
\label{Spelec}
\end{figure}
As a summary, we depict in Figure \ref{Spelec} (m)
the magnetic Type IIA limit.

\subsection{Dynamical Supersymmetry Breaking Model}

In \cite{DSB} an \MT fivebrane realization of
Izawa-Yanagida-Intriligator-Thomas (IYIT) model
of dynamical supersymmetry breaking was proposed.
IYIT model is a supersymmetric $Sp(N_c)$ QCD with
$N_f=N_c+1$ flavors coupled to gauge singlet chiral multiplets
$S_{ij}=-S_{ji}$ by the tree-level superpotential
\beq
W_{\it tree}=\lambda S_{ij}Q^iQ^j,
\label{WIYIT}
\eeq
where $\lambda$ is a coupling constant.
Since the equation $\partial W/\partial S_{ij}=0$ is in
conflict with the quantum modified constraint
${\rm Pf}Q^iQ^j=\Lambda^{4(N_c+1)}$
the model has no supersymmetric ground state.
However, an analysis of the K\"ahler potential of the singlet field
$S_{ij}$ at large values shows that there is a stable vacuum.

$W_{\it tree}$ in (\ref{WIYIT})
looks like the tree-level superpotential
of the magnetic theory considered above.
Therefore, the goal is to find a fivebrane configuration
that reduces to the Type IIA brane configuration as depicted
in Figure \ref{Spelec} (m) with $\tilNc$ replaced by
$N_c$, which exhibits the correct behaviour of the
gauge coupling constant at an energy much larger than
$\Lambda$.

The proposal of \cite{DSB} is essentially
that the fivebrane configuration is
given by that of the ``electric dual'', which is
$Sp(N_e)$ SQCD,
$N_e=N_f-N_c-2$, whose dynamical scale $\Lambda_e$ is
related to the scale $\Lambda$ of the original model
by $\Lambda^{3(N_c+1)-N_f}\Lambda_e^{3(N_e+1)-N_f}=\lambda^{-N_f}$.
Actually, in the present case electric-magnetic duality does not hold
since $N_e=(N_c+1)-N_c-2=-1$.
However, the equation (\ref{Spconf})
does make sense even if we put $N_c=-1$ and we can use it
to define the holomorphic asymptotic condition of the
fivebrane, thus defining a supersymmetric field
theory in four-dimension.
This proposal was shown to be consistent with the field theory
results by observing an agreement of the moduli space of
supersymmetric vacua of some perturbed models,
including the model perturbed by a linear term in $S$,
\beq
W_{\it tree}=\lambda SQQ+mS,
\eeq
which
has supersymmetric vacua when $(m/\lambda)^{N_f}=\Lambda^{2N_f}$.
Here we provide a more elementary evidence by showing
that the configurations
have the required Type IIA limit.

The configuration of the model perturbed by the linear term
$mS$
is in the eleven-dimensional space-time $\R^4\times M^7/\Z_2$
as in the case of $Sp(N_c)$ SQCD but now the Taub-NUT space
is deformed as $yx=(v^2+m^2)^{N_f}$.
The supersymmetric configuration of the model with
$(m/\lambda)^{N_f}=\Lambda^{2N_f}$ given in \cite{DSB}
is the configuration
for the ``electric dual'' (\ref{Spconf})
where $N_c$ and $\Lambda$ there
is replaced by $N_e=-1$ and $\Lambda_e$.
Using the relation between $\Lambda_e$ and $\Lambda$,
it is given by
\beq
C^{\prime}~
\left\{
\begin{array}{l}
y=1\\
v=0,
\end{array}
\right.
~~~~~
C~\left\{
\begin{array}{l}
x=\lambda^{2(N_c+1)}\Lambda^{4(N_c+1)}\\
w=0.
\end{array}
\right.
\label{IYITconf}
\eeq
(There is no $\CP^1$ component of the fivebrane.)
The component $C$ intersects with the $\Z_2$ fixed plane
$v=w=0$ at $x=\lambda^{2N_f}\Lambda^{4N_f}$
while the other component
$C^{\prime}$ intersects at $x=m^{2N_f}$.
Therefore, the configuration is consistent only if
$(m/\lambda)^{2N_f}=\Lambda^{4N_f}$, which is (almost) the same
as the condition for unbroken supersymmetry in field theory.
\footnote{Actually, the configuration (\ref{IYITconf})
can be deformed to a single component curve given by
$vw=\lambda\Lambda^2\sigma$ and $x=\lambda^{2(N_c+1)}\Lambda^{4(N_c+1)}$,
corresponding to the one-dimensional modulus space
in field theory where $\sigma$ corresponds to the eigenvalue of
the singlet $S$.
However, we may restrict our attention to the $\sigma=0$ case
(\ref{IYITconf}), since we are interested in the energy
much larger than $\Lambda$ and $S$.}

As before, we consider taking the limit $\Lambda\ll E$
for a fixed energy scale $E$ and look at the configuration
(\ref{IYITconf})
with respect to some fixed length scales in the $v$ and $w$ direction.
As in the magnetic limit of SQCD,
\begin{figure}[htb]
\begin{center}
\epsfxsize=4.5in\leavevmode\epsfbox{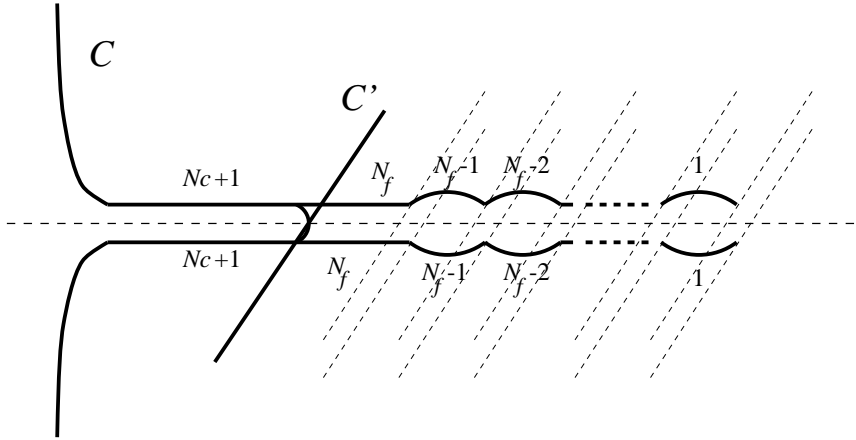}
\end{center}
\caption{The $M$ Theory Configuration for the Perturbed
IYIT Model in the
  Small $\Lambda$ Limit}
\label{IYITfig}
\end{figure}
the component $C$ is sent to the left
and wraps on the $\CP^1$ cycles as depicted in Figure \ref{IYITfig}.
In the limit $R\ll\elel$, this reduces to a Type IIA brane
configuration that obviously corresponds to IYIT model perturbed by
a (very small but finite) linear term in $S$.
If we force the pairs of D6-branes to approach each other at
the orientifold plane, and then separate them in the $x^6$
direction, we will get the Type IIA configuration that looks like
Figure \ref{Spelec} (m) with $\tilNc$ replaced by $N_c$.
Note that the curve is bent and is no longer holomorphic,
and therefore it has some non-zero energy density.

Unlike in the magnetic dual of SQCD, there is no bending of the component
$C^{\prime}$, and it appears that there is no natural way of setting
 the scale in the $w$ direction to measure the gauge coupling constant.
However, we can use a knowledge about the perturbed models considered
in \cite{DSB}.
Since the $w$ value of some characteristic point in such perturbed
models is identified with the gauge singlet
$S_{ij}$ which give a mass $\lambda S_{ij}$
to the quarks, the scale in the $w$ direction is set by $w\sim
\lambda^{-1}E$. Then, the scale in the $v$ direction is set by
$v\sim \lambda E^2$. Thus, the gauge coupling is measured as
$\e^{-2/g^2}=(\Lambda/E)^{4N_c+4}$ which is the
correct behaviour.

\section{Orthogonal Groups}

Here we consider the orthogonal gauge groups.
The \MT configuration for $N=1$ SQCD with even flavors
can be obtained by rotating the $N=2$ configuration
constructed in \cite{LLL,H2}.
We find two Type IIA limits corrersponding to the elecetric
and magnetic description of the theories.

\subsection{The Fivebrane Configuration}

The fivebrane configuration for $N=1$ $SO(N_c)$ SQCD
with $N_f$ flavors ($N_f$ chiral multiplets in the vector
representation) can be obtained by rotating the $N=2$
configuration in the standard way \cite{HOO} if
the flavor $N_f$ is even ($N_f$ here means $2N_f$ of ref.\cite{H2}).
We first consider the range $N_f\geq N_c-1$.
We present the result separately
for even and odd $N_c$.

\noindent
\underline{\sl $N_c$ Even, $N_f$ Even}

The space-time in which the fivebrane is embedded is
the $\Z_2$ quotient of the resolved
$A_{N_f-1}$ Taub-NUT space $yx=v^{N_f}$ where $\Z_2$ acts as
$y\to y, x\to x$ and $v\to -v$.
The $\Z_2$ fixes the $\CP^1$ cycles $C_{2i}$
point-wisely, but acts on $C_{2i+1}$ as the $\pi$-rotation
around the intersection points with $C_{2i}$ and $C_{2i+2}$.

The fivebrane consists of infinite components in $x^7=0$
\beq
C^{\prime}~
\left\{
\begin{array}{l}
w^2y=\prod_{i=1}^{N_c/2}(w^2-M_i^2)\\
v=0
\end{array}
\right.
~~~~~
C~\left\{
\begin{array}{l}
v^2x=\Lambda^{-(3(N_c-2)-N_f)}v^{N_c}\\
w=0,
\end{array}
\right.
\label{SOevenconf}
\eeq
and $\CP^1$ components wrapping the cycles $C_i$ with multiplicity
$N_c$ for $i=1,\ldots, N_f-N_c+2$ and $N_f-i+2$ for $i=N_f-N_c+3,\ldots,
N_f-1$. 
Note that $C$ includes a component described by $v^2=0$, the fivebrane
wrapping twice on the $x$-axis. Also, the component $C^{\prime}$
at large $y$ looks almost like the fivebrane wrapping
twice on the $y$-axis in $x^{7,8,9}=0$.
These regions --- large $|x^6|$ with $x^{4,5,7,8,9}=0$ --- 
correspond in the Type IIA limit to
the O4${}^+$-plane ($Sp$-type, trivial RR Wilson line).
The main component of $C$, $x=\Lambda^{-(3(N_c-2)-N_f)}v^{N_c-2}$
(which we denote again by $C$),
intersects transversely
with the $\CP^1$ cycle $C_{N_f-N_c+2}$.
In addition, the fivebrane wraps odd number of times
 on the cycles $C_{2i+1}$ (on which $\Z_2$ acts as the $\pi$-rotation)
for $N_f-N_c+3\leq 2i+1\leq N_f-1$.  
 In order to avoid
{\it t}-configurations, at least one pair of components
wrapping $C_{2i}$ must be at $x^{7,8,9}=0$ for each $2i$ in
$N_f-N_c+2\leq 2i<N_f-1$.

\begin{figure}[htb]
\begin{center}
\epsfxsize=5.8in\leavevmode\epsfbox{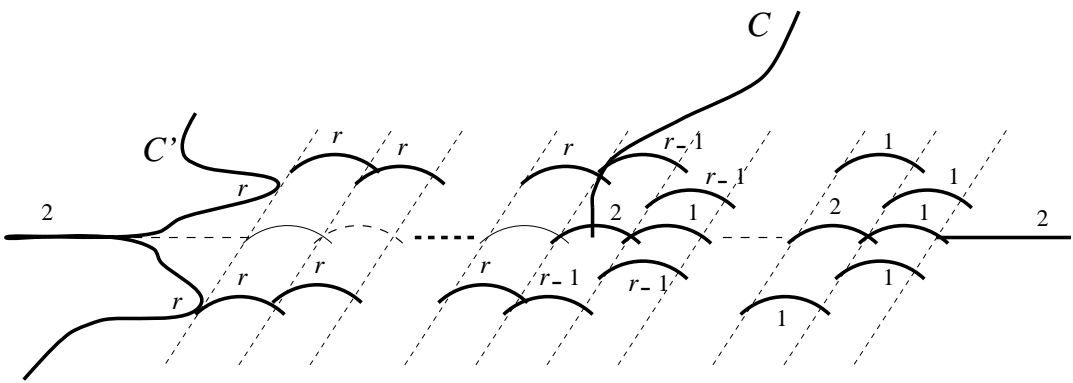}
\end{center}
\caption{The $M$ Theory Configuration of $SO(2r)$ SQCD.}
{\small There are $N_f$ sixbranes. The dashed horizontal lines
and curves are point-wisely $\Z_2$ fixed while
$\Z_2$ acts on the thin curves as the $\pi$-rotation.}
\label{SOevenfig}
\end{figure}

The number of $\Z_2$ invariant
deformations of the curve subject to this constraint counts as
\beq
{N_c\over 2}+2\left[
(N_f-N_c+1){N_c\over 2}
+2{({N_c\over 2}-1){N_c\over 2}\over 2}\right]
=N_cN_f-{N_c(N_c-1)\over 2}
\eeq
which agrees with the dimension of the moduli space of vacua.
The parameters $\pm M_i$ are interpreted as the eigenvalues of
$J_{ij}M^{jk}$ where $M^{ij}=Q^i_aQ^j_a$ is the vev of the
meson matrix and $J_{ij}$ is the $Sp(N_f/2)$ invariant
skew-symmetric form. Note that the eigenvalues of $JM$ come in pairs
since ${}^t(JM)=-J^{-1}(JM)J$.

\noindent
\underline{\sl $N_c$ Odd, $N_f$ Even}

The space-time in which the fivebrane is embedded is
the $\Z_2$ quotient of the resolved
$A_{N_f-1}$ Taub-NUT space $yx=v^{N_f}$ where $\Z_2$ acts as
$y\to -y, x\to -x$ and $v\to -v$.
The $\Z_2$ fixes the $\CP^1$ cycles $C_{2i+1}$
point-wisely, but acts on $C_{2i}$ as the $\pi$-rotation
around the intersection points with $C_{2i-1}$ and $C_{2i+1}$.

The fivebrane consists of infinite components in $x^7=0$
\beq
C^{\prime}~
\left\{
\begin{array}{l}
wy=\prod_{i=1}^{[N_c/2]}(w^2-M_i^2)\\
v=0
\end{array}
\right.
~~~~~
C~\left\{
\begin{array}{l}
vx=\Lambda^{-(3(N_c-2)-N_f)}v^{N_c-1}\\
w=0,
\end{array}
\right.
\label{SOoddconf}
\eeq
and $\CP^1$ components wrapping the cycles $C_i$ with multiplicity
$N_c\pm 1$ for $i=1,\ldots,N_f-N_c+2$,
and $N_f-i+2\pm 1$ for $i=N_f-N_c+3,\ldots,
N_f-1$ where upper case ($+1$) is for odd $i$ and lower case ($-1$)
is for even $i$. 
Note that $C$ includes a component described by $v=0$, the fivebrane
wrapping once on the $x$-axis. Also, the component $C^{\prime}$
at large $y$ looks almost like the fivebrane wrapping
once on the $y$-axis.
Noting that $\Z_2$ acts on the $y$ and $x$ axis at $x^{7,8,9}=0$
as the $\pi$-rotation around $y=0$ and $x=0$, we see that
these regions
--- large $|x^6|$ with $x^{4,5,7,8,9}=0$ --- 
correspond in the Type IIA limit to
the $\tilO$-plane ($Sp$-type, non-trivial RR Wilson line).
The main component of $C$, $x=\Lambda^{-(3(N_c-2)-N_f)}v^{N_c-2}$
(which we denote again by $C$),
intersects transversely
with the $\CP^1$ cycle $C_{N_f-N_c+2}$.
In addition, the fivebrane wraps odd number of times
 on the cycles $C_{2i}$ (on which $\Z_2$ acts as the $\pi$-rotation)
for $N_f-N_c+3\leq 2i\leq N_f-1$
and also on the $x$-axis.
 In order to avoid
a {\it t}-configuration,
at least one pair of components
wrapping $C_{2i+1}$ must be at $x^{7,8,9}=0$ for each $2i+1$ in
$N_f-N_c+2\leq 2i+1\leq N_f-1$.

\begin{figure}[htb]
\begin{center}
\epsfxsize=5.8in\leavevmode\epsfbox{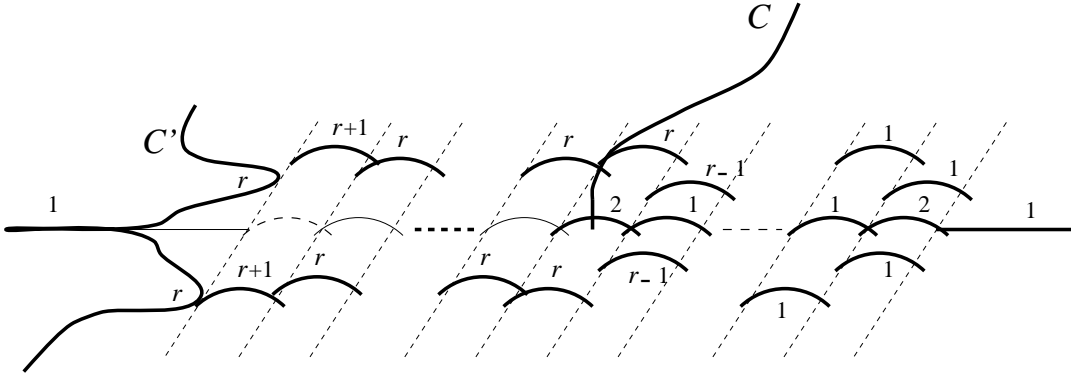}
\end{center}
\caption{The $M$ Theory Configuration of $SO(2r+1)$ SQCD.}
{\small There are $N_f$ sixbranes. The thin horizontal lines
and curves are $\pi$-rotated by $\Z_2$
while the dashed curves are point-wisely $\Z_2$ fixed. Note the
difference from $SO(2r)$ case.}
\label{SOoddfig}
\end{figure}

The number of $\Z_2$ invariant deformations of the curve
subject to this constraint counts as
\beq
{N_c-1\over 2}+2\left[
\mbox{${N_f-N_c+1\over 2}{N_c+1\over 2}
+{N_f-N_c+1\over 2}{N_c-1\over 2}+{N_c-1\over 2}$}
+2{{N_c-3\over 2}{N_c-1\over 2}\over 2}\right]
=N_cN_f-{N_c(N_c-1)\over 2}
\eeq
which agrees with the dimension of the moduli space of vacua.
The parameters $\pm M_i$ are interpreted as the eigenvalues of
$J_{ij}M^{jk}$.

\subsection{The Electric and Magnetic Limits}

We examine the weakly coupled Type IIA limit of the \MT configuration
described above.
In particular, we will find two limits which realizes the electric
and magnetic theories of \cite{ISp}.
We explicitly write down the procedure for even $N_c$, but not
for odd $N_c$ because the result is essentially the same and we only
need a little more care.

\subsection*{\sl The Electric Limit}

Let us consider the limit 
$\Lambda^{3(N_c-2)-N_f}\ll E^{3(N_c-2)-N_f}$
for a fixed energy scale $E$ and
look at the configuration
with respect to the scale set by $v\sim E$, $w\sim E^2$.
The main component of $C$ is sent to the right
and wraps on the $\CP^1$ cycles
$C_{N_f-N_c+3},\ldots, C_{N_f-1}$ with multiplicities
$1,2,\ldots, N_c-3$.
It also wraps on the $x$-axis with mulitplicity $N_c-2$
and then blows up in the $v$-direction at
$y(E)=\Lambda^{3(N_c-2)-N_f}E^{N_f-(N_c-2)}$.
The component $C^{\prime}$ wraps on the $y$-axis and blows up in the
$w$-direction at $y^{\prime}(E)=E^{2(N_c-2)}$.
The two regions with large $v\gg E$ and large $w\gg E^2$
are interpreted as the NS and NS${}^{\prime}$5-branes
in the Type IIA limit.
Between them, the fivebrane wraps
$N_c$ times on the eleventh direction.
In the weakly coupled Type IIA limit, these are identified as
the $N_c$ D4-branes stretched between the NS${}^{\prime}$ and the
NS5-branes.
The $\Z_2$ fixed cycle $C_{2i}$ at $x^{7,8,9}=0$ is
identified as O4${}^-$-plane
whereas the $\Z_2$ invariant but non-fixed cycle $C_{2i+1}$
at $x^{7,8,9}=0$
is identified as the O4${}^0$-plane.
Namely, the region of O4-plane
sandwiched between the two NS5-branes is
basically of $SO$-type.
Therefore the $N_c$ D4-branes yields $SO(N_c)$ gauge symmetry.

The gauge coupling constant $g(E)$ at the energy $E$
can be read, as in the $SU(N_c)$ case, as
\beq
\e^{-1/g^2(E)}={y(E)\over y^{\prime}(E)}
=\left({\Lambda\over E}\right)^{3(N_c-2)-N_f}.
\eeq

As a summary, we depict in Figure \ref{SOIIA} (e)
the electric Type IIA limit.
Note that the dimension of the moduli space can be counted also
by applying the ``s-rule'' found in \cite{H2} to
this configuration.

\subsection*{\sl The Magnetic Limit}

Let us next consider the opposite limit
$\Lambda^{3(N_c-2)-N_f}\to \infty$.
The component $C$ is sent to the left and wraps on the $\CP^1$ cycles
$C_{1},\ldots, C_{N_f-N_c+1}$ with multiplicities
$N_f-N_c+1,\ldots,2,1$.
It also wraps $N_f-N_c+2$ times
on the $y$-axis and starts blowing up in the $v$-direction
at some value of $y$.
This value of $y$ is much larger than
the value where the component $C^{\prime}$ starts blowing up in the
$w$-direction with respect to some scale which will be specified
momentarily.
Between the two regions with large $v$ and $w$,
the fivebrane wraps $\tilNc=N_f-N_c+4$ times
on the eleventh direction.
In the Type IIA limit, it is
interpreted as $\tilNc$ degenerate D4-branes.
Since these are on top of the O4${}^-$-plane,
we obtain $SO(\tilNc)$ gauge symmetry.
We can measure the gauge coupling at an energy $E$
by setting
the corresponding length scales in the $v$ and $w$ directions.
As in $SU(N_c)$ case, these length scales are set by
$w\sim \mu E$ and $v\sim\mu^{-1}E^2$ where $\mu$ is
a constant mass scale,
and the $SO(\tilNc)$ gauge coupling $\widetilde{g}(E)$
is measured as
\beq
\e^{-1/\widetilde{g}^2(E)}=
{\mu^{N_f}\Lambda^{-3(\tilNc-2)+N_f}\over E^{3(\tilNc-2)-N_f}}.
\eeq
This yields the standard relation
$\Lambda^{3(\tilNc-2)-N_f}\tiLambda^{3(\tilNc-2)-N_f}
=\mu^{N_f}$
between the dynamical scales of
the electric and magnetic theories \cite{IS}.

On the right of the region of $C^{\prime}$ with large $w\gg\mu E$
(which is identified as the NS${}^{\prime}$5-brane),
we have components of the fivebrane wrapping
several times on the eleventh direction.
The fivebrane wraps $N_f+2$-times in the region sandwiched between
the NS${}^{\prime}$5-brane and the left-most D6-brane,
while it wraps $i+2$-times on the $\CP^1$ cycle $C_{N_f-i}$.
In the weakly coupled Type IIA limit $R\ll \elel$,
two of the fivebranes in the region sandwiched between
the NS${}^{\prime}$5-brane and the left-most D6-brane
are interpreted as providing a
charge $+2$ to the orientifold fixed plane
(consituting the O4${}^+$-plane).
Also,
a pair of fivebranes wrapping each of the $\Z_2$ fixed cycles
(namely, the $\CP^1$ cycles $C_{2i}$ and the $x$-axis
at $x^{7,8,9}=0$) is interpreted as the O4${}^+$-plane.
Similarly, a single fivebrane wrapped on each of the $\Z_2$
invariant cycles $C_{2i+1}$ at $x^{7,8,9}=0$
 (on which $\Z_2$ acts as the
$\pi$-rotation) is interpreted as the $\tilO$-plane.
The rest of the fivebrane wraps $N_f$-times
in the region sandwiched between
the NS${}^{\prime}$5-brane and the left-most D6-brane,
while it wraps $2[(i+1)/2]$-times
on the cycles $C_{N_f-i}$.
These are interpreted as the D4-branes
which represent the degrees of freedom
corresponding to
the gauge singlet meson field of the magnetic theory.
Indeed, the number of possible deformations of these D4-branes
is
\beq
{N_f\over 2}+2\sum_{i=1}^{N_f-1}[(i+1)/2]
={N_f(N_f+1)\over 2},
\eeq
which is equal to the number of independent components
of the singlet meson field $\tilde{M}^{ij}=\tilde{M}^{ji}$.

As a summary, we depict in Figure \ref{SOIIA} (m)
the magnetic Type IIA limit.
The O4-planes that extend to the left and right infinity
are O4${}^+$-plane.
O4${}^+$
and $\tilO$ appears alternately since D6-brane is a unit magnetic monopole
for the RR $U(1)$ gauge field.

\subsection*{\sl Odd $N_c$}

\begin{figure}[htb]
\begin{center}
\epsfxsize=6.2in\leavevmode\epsfbox{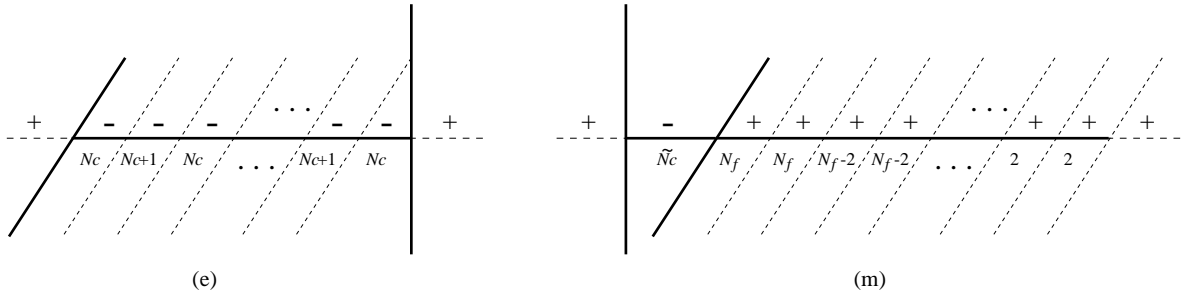}
\end{center}
\caption{The Electric (e) and Magnetic (m) Type IIA Limits
(Orthogonal Gauge Groups) There are $N_f$ D6-branes in each.}
\label{SOIIA}
\end{figure}

We can consider the Type IIA limit for odd $N_c$ in a similar way.
It turns out that the result is the same:
we can find the electric and magnetic limits which
are given exactly as in Figure
\ref{SOIIA} (e) and (m), where $N_c$ and $\tilNc$
are odd now.
The O4-planes that extend to the left and right infinity
are $\tilO$-plane in the present case.
In the magnetic configuration, $\tilO$
and O4${}^+$ again appears alternately as we cross D6-branes.

\subsection{$N_f=N_c-4,N_c-3$: Confining Phase}

In the rest of the paper, we consider the remaining cases
$0<N_f<N_c-2$. We again restrict our attention to the case where
$N_f$ is even.

\noindent
\underline{$N_f<N_c-4$}

For $N_f<N_c-4$, there is no fivebrane configuration
that lifts the electric Type IIA configuration
(Figure \ref{SOIIA} (e)), as can be seen as follows.
The asymptotics at large $v$ and large $x$ requires that the
fivebrane has a component $C$ described by (\ref{SOevenconf})
or (\ref{SOoddconf}). However, the equation for $C$ implies
$yv^{N_c-2-N_f}=\Lambda^{3(N_c-1)-N_f}$
where the power of $v$ is at least 4 ($N_c$ even) or 3 ($N_c$ odd)
when $N_f<N_c-4$.
Since $v^2=0$ ($N_c$ even) or $v=0$ ($N_c$ odd) at large $y$
represents the $Sp$-type O4-plane,
this shows that there are extra D4-branes on the left of the
NS${}^{\prime}$5-brane, which is not the property of the
Type IIA configuration.
The absence of \MT configuration corresponds to the absence of
(supersymmetric) stable vacua due to the generation
of Affleck-Dine-Seiberg superpotential.

\noindent
\underline{$N_f=N_c-4$} {\sl ($N_c$ Even)}

The fivebrane consists of infinite components at $x^7=0$
\beq
C^{\prime}~
\left\{
\begin{array}{l}
y=w^2\prod_{i=1}^{N_f/2}(w^2-M_i^2)\\
v=0
\end{array}
\right.
~~~~~
C~\left\{
\begin{array}{l}
v^2x=\Lambda^{-(3(N_c-2)-N_f)}v^{N_c}\\
w=0,
\end{array}
\right.
\label{SOevenspecial}
\eeq
and $\CP^1$ components wrapping $C_1,\ldots,C_{N_f-1}$
with multiplicity $N_c-3,N_c-4,\ldots,4,3$.
Again $C$ includes a component wrapping twice on the $x$-axis.
The strong coupling limit $\Lambda\to\infty$ of this configuration
leads to a configuration which looks like
the magnetic configuration of Figure \ref{SOIIA} (m)
extrapolated to $N_f=N_c-4$, namely, $\tilNc=0$.
We see that there is no gauge symmetry in this configuration
but there are degrees of freedom as many as
$N_f(N_f+1)/2$ chiral superfields.
This is consistent with the field theory knowledge \cite{ISp}
that the theory confines and becomes a theory of
composite meson field $M^{ij}=Q^iQ^j$.

\noindent
\underline{$N_f=N_c-3$} {\sl ($N_c$ Odd)}

The fivebrane consists of infinite components at $x^7=0$
\beq
C^{\prime}~
\left\{
\begin{array}{l}
y=w\prod_{i=1}^{N_f/2}(w^2-M_i^2)\\
v=0
\end{array}
\right.
~~~~~
C~\left\{
\begin{array}{l}
vx=\Lambda^{-(3(N_c-2)-N_f)}v^{N_c-1}\\
w=0,
\end{array}
\right.
\label{SOoddspecial}
\eeq
and $\CP^1$ components wrapping $C_i$
with multiplicity $N_c-1-i\pm 1$ where $+1$ is for odd $i$
and $-1$ is for even $i$.
$C$ includes a component wrapping once over the $x$-axis.
The strong coupling limit $\Lambda\to\infty$ of this configuration
leads to a configuration which looks like
the magnetic configuration of Figure \ref{SOIIA} (m)
extrapolated to $N_f=N_c-3$, namely, $\tilNc=1$.
We see that there is no gauge symmetry in this configuration
but there are degrees of freedom as many as
$N_f(N_f+1)/2$ chiral superfields.
Also, there are $N_f$ modes
associated with the open string stretched between
the single D4-brane on top of O4-plane
and the $N_f$ D4-branes ending on the NS${}^{\prime}$5-brane.
The mass of these modes are given by $M_i$
which are identified as the meson eigenvalues.
This is consistent with the field theory knowledge \cite{ISp}
that the theory confines and becomes a theory of
composite meson field $M^{ij}$ and composite $N_f$ fields
$q_i$ (glueballs or exotics) coupled via a superpotential
of the form $M^{ij}q_iq_j$.

\subsection{$N_f=N_c-2$: Coulomb Phase}

The fivebrane for the case $N_f=N_c-2$ ($N_c$ even)
consists of infinite components at $x^7=0$
\beq
C^{\prime}~
\left\{
\begin{array}{l}
w^2y=w^2\prod_{i=1}^{N_f/2}(w^2-M_i^2)\\
v=0
\end{array}
\right.
~~~~~
C~\left\{
\begin{array}{l}
v^2x=\Lambda^{-2(N_c-2)}v^{N_c}\\
w=0,
\end{array}
\right.
\label{SOevenCou}
\eeq
and $\CP^1$ components wrapping $C_1,\ldots,C_{N_f-1}$
with multiplicity $N_c-1,N_c-2,\ldots,4,3$.
$C$ includes a component wrapping twice on the $x$-axis,
and $C^{\prime}$ includes a component wrapping twice
on the $y$-axis.
The main component of $C$ is decribed by
$y=\Lambda^{2(N_c-2)}$
and intersects transversely with the $y$-axis.
The main component of $C^{\prime}$ also intersects transversely
with the $y$-axis at $y=\prod_i(-M_i^2)$.
The $y$-axis is a $\Z_2$ fixed plane but the intersection
with a fivebrane is allowed
because it is screened by a pair of fivebranes (see \cite{H2}).

In the weak coupling region, all $M_i\gg \Lambda^2$,
$C^{\prime}$ is on the left of $C$ in a neighborhood of
the $y$-axis. 
In the strong coupling region, all $M_i\ll\Lambda^2$,
$C$ is on the left of $C^{\prime}$ in a neighborhood
of the $y$-axis, just as in the
magnetic configuration (Figure \ref{SOIIA} (m)).
For both cases,
in the weakly coupled Type IIA limit,
the region of the $y$-axis sandwiched
between $C$ and $C^{\prime}$ is identified as O4${}^-$-plane
with a pair of D4-branes. Namely, we see the $SO(2)$ gauge symmetry
in both weak and strong coupling regions.
In the strong coupling region, there are $N_f$ electrons
transforming in the vector representation of $SO(2)$
with mass given by $M_i$, as the megnetic configuration shows.
These two $SO(2)$ gauge theories, electric and magnetic
$SO(2)$s, are continuously connected
by changing $M_i$'s because the intersection point
$y=\prod_i(-M_i^2)$ of $C^{\prime}$
can go around the intersection point
$y=\Lambda^{2(N_c-2)}$ of $C$.
Something interesting happens when these intersection points
collide, $\prod_{i=1}^{N_f/2}(-M_i^2)=\Lambda^{2(N_c-2)}$.

These are qualitatively the same as
what we know for field theory.
However, we know from field theory considerations more than
just the qualitative features.
For example, the exact effective $SO(2)$ gauge coupling $g_{\it eff}$
has been determined by finding the Seiberg-Witten curve \cite{ISp}.
In the brane picture, it is not obvious how to compute
it. We can only tell that
$1/g^2\sim {\rm log}(\prod_iM_i^2/\Lambda^{2(N_c-2)})$
in the weak coupling region, which is roughly correct but not exact.
The curve is completely degenerate and there is no way to
make it a smooth genus one Riemann surface without changing the
asymptotic behaviour, as can be seen by noting
the existence of a meromorphis function with a single simple pole.
The effective gauge coupling constant should be determined in a
different way than using the Seiberg-Witten type curve.
Also, the field theory argument \cite{ISp} shows
that there is a dyon which becomes massless at the colliding point
$\prod_{i=1}^{N_f/2}(-M_i^2)=\Lambda^{2(N_c-2)}$.
It is not clear whether such a dyon state exists in the brane
picture.
It is challenging to solve these problems.
It might be related to something important in \MT on
$\R^5/\Z_2$ orbifold.

\vfill
\noindent
{\bf Acknowledgement}

I would like to thank C. Csaki,
J. de Boer, A. Giveon, H. Murayama, H. Ooguri, Y. Oz,
S.-J. Rey, M. Schmaltz, M. Strassler, R. Sundrum
and C. Vafa
for discussions.
I thank Institute for Theoretical Physics at Santa Barbara
for hospitality.

This research is supported in part by
NSF grant PHY-95-14797
and
DOE grant DE-AC03-76SF00098,
and also by NSF grant PHY94-07194.

\appendix{Vacuum Structure of $N=2$ SQCD Perturbed to $N=1$}

In the previous work \cite{HOO}, $N=2$ $SU(N_c)$ SQCD with $N_f$ flavors
perturbed by a mass term for the adjoint chiral multiplet
was studied both from field theory method and from brane description.
For $N_f>N_c+1$,
there was a gap between the results of the two methods concerning
the vacua with meson matrix of rank $<N_f$.
In this appendix, we correct an error in the
field theory treatment of \cite{HOO}
and present a better argument showing that the brane result
was correct.

By integrating out the adjoint chiral multiplet of mass $\mua$,
the system can be considered as $N=1$ SQCD perturbed by the
chiral symmetry breaking superpotential
\beq
{\sl \Delta}W={1\over 2\mua}\left[\,
\Tr(\tilQ Q\tilQ Q)-{1\over N_c}(\Tr \tilQ Q)^2\right]
\label{treep}
\eeq
at the cut-off scale $\mua$.
As far as the gauge group $SU(N_c)$ is broken at high energy
by a large vev of $Q$ or $\tilQ$, one can use this to analyze the
space of vacua. However, once $Q$ and $\tilQ$ become smaller than
$\Lambda$ (as it turns out, all branch of the moduli space is in or
connected to this region), the gauge coupling becomes strong
and it is no longer valid to use the elementary field.
For $N_c+1<N_f\leq 3N_c/2$, it is appropriate to use the magnetic
description of the theory which becomes weakly coupled at low energy.
Even for $3N_c/2<N_f<2N_c$,
although the theory without perturbation (\ref{treep}) flows to
a non-trivial fixed point, it turns out that the
moduli space of the theory with (\ref{treep})
is in or connected to a region where we can find
an appropriate description starting from the magnetic theory.
Thus, we consider the $SU(\tilNc)$ gauge theory with $N_f$ quarks
$q_i,\widetilde{q}^i$ (magnetic quarks) and
a gauge singlet meson $M$ ($N_f\times N_f$ matrix)
which are coupled by tree-level superpotential
\beq
W_{\it tree}
={1\over \mu}Mq\widetilde{q}+
{1\over 2\mua}\left[\, \Tr(M^2)-{1\over N_c}(\Tr M)^2\right].
\eeq

The error of the treatment in \cite{HOO} was that we assumed
following \cite{EGKRS} that there exists an exact superpotential that
applies for any rank of $M$ (= the sum of $W_{\it tree}$
and an Affleck-Dine-Seiberg type potential).
However, this potential is not holomorphic at $\det M=0$
and the analysis is not valid for ${\rm rank}M<N_f$.
In fact, the analysis should depend on the
rank of $M$ since the low energy theory depends drastically on the
number of massless quarks which is determined by rank$M$.

If the rank of $M$ is $N_f$,
all the magnetic quarks are massive and the low energy theory
is the pure $SU(\tilNc)$ super-Yang-Mills theory with the dynamical scale
$\Lambda_L$ given by $\Lambda_L^{3\tilNc}=\det(M/\mu)
\tiLambda^{3\tilNc-N_f}$.
Thus, the theory in this branch is described by the exact superpotential
\beq
W_{\it eff}=\tilNc \Lambda_L^3+
{1\over 2\mua}\left[\, \Tr(M^2)-{1\over N_c}(\Tr M)^2\right].
\eeq
Extremizing $W_{\it eff}$, we obtain the maximal rank solutions
which were obtained in \cite{HOO} both
from field theory and from brane analysis.
Indeed, the solution is of order $M\sim
(\Lambda^{3N_c-N_f}/\mua^{N_f-N_c})^{1/(2N_c-N_f)}\ll\Lambda^2$
for $\mua\gg\Lambda$, and is inaccessible from the electric description.

If the rank $\ell$ of $M$ is smaller $\ell<N_f$,
$N_f^{\prime}=N_f-\ell$ of the quarks remain massless.
Let us decompose the meson matrix into blocks of size
$\ell$ and $N_f^{\prime}$ as
\beq
M\,=\left(
\begin{array}{cc}
M_H&M_{12}\\
M_{21}&M_L
\end{array}
\right)
\eeq
where ${\rm rank}\langle M_H\rangle=\ell$, 
$\langle M_{12}\rangle=\langle M_{21}\rangle=\langle M_L\rangle=0$.
By integrating out the $\ell$ massive quarks with mass matrix $M_H/\mu$,
we obtain the low energy theory which is $SU(\tilNc)$ SQCD with
$N_f^{\prime}=N_f-\ell$ quarks $q_{Li},\widetilde{q}_{L}^i$
with the dynamical scale $\Lambda_L$ and the tree-level
superpotential $W_L$ given by
\beqa
&&\Lambda_L^{3\tilNc-N_f^{\prime}}=\det\left({M_H\over \mu}\right)
\tiLambda^{3\tilNc-N_f},
\label{matcha}\\
W_L&=&{1\over\mu}M_Lq_L\widetilde{q}_L
+{1\over 2\mua}\left[\, \Tr(M^2)-{1\over N_c}(\Tr M)^2\right].
\eeqa

For $N_c<\ell<N_f$, since the flavor $N_f^{\prime}$ of the low energy
theory is smaller than the color $\tilNc$, Affleck-Dine-Seiberg
superpotential is generated
\beq
W_{\it eff}\,=\,W_L
+(\tilNc-N_f^{\prime})\left({\Lambda_L^{3\tilNc-N_f^{\prime}}\over
\det q_L\widetilde{q}_L}\right)^{1/(\tilNc-N_f^{\prime})}.
\eeq
It is easy to see that there is no solution to
$\partial W_{\it eff}=0$ which satisfies $M_L=0$.
Hence there is no vacuum with $N_c<{\rm rank}M<N_f$.

For $\ell=N_c$, the low energy theory is $SU(\tilNc)$ with flavor
$N_f^{\prime}=\tilNc$.
At energies far below $\Lambda_L$, this theory is described by composite
mesons $N_L=q_L\widetilde{q}_L$ and baryons $b_L=\det q_L$,
$\widetilde{b}_L=\det\widetilde{q}_L$ which are subject to
the quantum modified constraint
$\det N_L -b_L\widetilde{b}_L=\Lambda_L^{2\tilNc}$.
Thus, the effective superpotential is
\beq
W_{\it eff}={1\over\mu}{\rm tr}(M_LN_L)+
{1\over 2\mua}\left[\, \Tr(M^2)-{1\over N_c}(\Tr M)^2\right]
+X\left(\det N_L -b_L\widetilde{b}_L-\Lambda_L^{2\tilNc}\right).
\eeq
The equation $\partial W_{\it eff}=0$ together with
$M_L=M_{12}=M_{21}=0$ requires
\beqa
&&N_L={\mu\over N_c\mua}\Tr M_H {\bf 1}_{\tilNc}\\
&&M_H-{1\over N_c}\Tr M_H=0\\
&&\det N_L -b_L\widetilde{b}_L=\Lambda_L^{2\tilNc}\label{qmc}
\eeqa
The first and second equations implies
$M_H=m{\bf 1}_{N_c}$ and $N_L=(\mu m/\mua){\bf 1}_{\tilNc}$
for some $m$. Then the quantum modified constraint (\ref{qmc}) of
the low energy theory yields
\beq
(m/\mua)^{\tilNc}-b_L\widetilde{b}_L/\mu^{\tilNc}
=\Lambda_L^{2\tilNc}/\mu^{\tilNc}.
\eeq
By using the relation between the electric and magnetic baryon
operators $b_L=b_{N_c+1,\ldots,N_f}=C^{-1} B^{1,\ldots,N_c}$ and
$\widetilde{b}_L=\widetilde{b}^{N_c+1,\ldots,N_f}
=C^{-1} \widetilde{B}_{1,\ldots,N_c}$,
$C=(-(-\mu)^{-\tilNc}\Lambda^{3N_c-N_f})^{1/2}$,
together with the scale matching equation (\ref{matcha}),
we obtain
\beq
m^{N_c}+B\widetilde{B}
=(-1)^{N_f-N_c}\left({m\over \mua}\right)^{N_f-N_c}\Lambda^{3N_c-N_f}.
\label{answer}
\eeq
This is nothing but the relation obtained in
\cite{HOO} from the brane description of the theory.

For $\ell=N_c-1$, the low energy theory is $SU(\tilNc)$ SQCD
with $N_f^{\prime}=\tilNc+1$ flavors.
At energies far below $\Lambda_L$, this theory is described by
composite mesons $N_L$ and baryons $b_L,\widetilde{b}_L$
and the superpotential is given by
\beq
W_{\it eff}={1\over\mu}{\rm tr}(M_LN_L)+
{1\over 2\mua}\left[\, \Tr(M^2)-{1\over N_c}(\Tr M)^2\right]
+\frac{1}{\Lambda_L^{2\tilNc-1}}
(\det N_L -b_LN_L\widetilde{b}_L).
\eeq
It is easy to see that there is no solution to
$\partial W_{\it eff}=0$ with $M_L=0$
and $\det M_H\ne 0$. Therefore, there is no vacuum with
${\rm rank}M=N_c-1$.

For $\ell<N_c-1$, the low energy theory is $SU(\tilNc)$ SQCD
with $N_f^{\prime}>\tilNc$. It is known that this theory exhibits
no generation of superpotential,
and the classical analysis for determining vacua is valid
as far as the moduli space extends to a region
where the gauge coupling stops running at high energy.
The vacuum equation
$\partial W_L=0$ is satisfied when $M_H=0$ (which implies $\ell=0$)
and $q_i\widetilde{q}^{j}=0$, and in fact either
$q_i$'s or $\widetilde{q}^j$'s can have large expectation values
of rank $\tilNc$.
From the equation $q_i\widetilde{q}^{j}=0$, we see that
$b_{i_1,\ldots,i_{\tilNc}}\widetilde{b}^{j_1,\ldots,j_{\tilNc}}=0$.
Thus, this branch is given by $M=0$,
$B^{i_1,\ldots,i_{N_c}}\widetilde{B}_{j_1,\ldots,j_{N_c}}=0$.
This is actually the $m\to 0$ limit of the branch already obtained above
(\ref{answer}).

One can also show that the brane tells the dimension of the moduli
space correctly.
For the maximal rank solutions, we have already seen in \cite{HOO}
that both field theory and brane analysis gives complex dimension
$2r(N_f-r)$.
For the other branch, the brane consists of $N_c(N_f-N_c)$ $\CP^1$
components and two factorized infinite components. If we include the
relative motion of the infinite components as one of its moduli
(which is still an unanswered problem), the complex dimension is
$2(N_fN_c-N_c^2+1)$.
On the field theory side, as we have seen above,
the solution can be made to the following form by the complexified
flavor symmetry $GL(N_f,\C)$:
\beqa
&&M=\left(\begin{array}{cc}
m{\bf 1}_{{}_{N_c}}&\\
&{\bf 0}_{{}_{N_f-N_c}}
\end{array}\right)
\\
&&B^{1,\ldots,N_c}=B,~~\mbox{other components}=0\\
&&\widetilde{B}_{1,\ldots,N_c}=\widetilde{B},
~~\mbox{other components}=0
\eeqa
where $m, B$ and $\widetilde{B}$ obey (\ref{answer}).
The unbroken subgroup of the flavor symmetry $GL(N_f,\C)$
at this solution is
$SL(N_c,\C)\times GL(N_f-N_c,\C)$.
Since the flavor invariants are parametrized by one complex modulus
due to the constraint (\ref{answer}), the complex dimension of the
moduli space is $1+N_f^2-(N_c^2-1+(N_f-N_c)^2)=
2(N_fN_c-N_c^2+1)$
which agrees with the result of the brane analysis.

\end{document}